\documentclass[10pt, conference, compsocconf]{IEEEtran}

\usepackage{subfigure}
\usepackage{graphicx}
\usepackage{amsmath}
\usepackage{bm}
\usepackage{url}
\usepackage{stmaryrd}
\usepackage{balance}
\usepackage{amssymb}
\usepackage{pgfplots}
\usepackage{pifont}
\usepackage[usenames,dvipsnames]{pstricks}
\usepackage{epsfig}
\usepackage{booktabs}
\usepackage{pgffor}
\usepackage{tikz}
\usepackage{multirow}

\usepackage{amsmath}

\newtheorem{theorem}{Theorem}

\newcommand{\mb}{\mathbf}

\newcommand{\problem}{\textsc{Tim}}
\newcommand{\problemknown}{\textsc{C-Tim}}
\newcommand{\problemunknown}{\textsc{J-Tim}}

\newcommand{\our}{\textsc{Tier}}
\newcommand{\ourknown}{\textsc{C-Tier}}
\newcommand{\ourunknown}{\textsc{J-Tier}}
\newcommand{\LTRandom}{\textsc{LT-random}}
\newcommand{\LTPageRank}{\textsc{LT-page rank}}
\newcommand{\LTGreedy}{\textsc{LT-greedy}}
\newcommand{\LTInDegree}{\textsc{LT-in degree}}

\newcommand{\gamecompeting}{\textsc{G-comp}}
\newcommand{\gamecollaborative}{\textsc{G-cpl}}
\newcommand{\gameindependent}{\textsc{G-indep}}

\newcommand{\diffusion}{\textsc{Tlt}}

\usepackage{algorithm}
\usepackage{algorithmic}

\begin{document}
\title{Intertwined Viral Marketing through Online Social Networks}

%
%

\author{
Jiawei Zhang$^\star$, Senzhang Wang$^\dagger$, Qianyi Zhan$^\ddagger$, Philip S. Yu$^\star$\\
      {$^\star$ University of Illinois at Chicago, Chicago, IL, USA}\\ 
      {$^\dagger$ Beihang University, Beijing, China}\\
       {$^\ddagger$ Nanjing University, Nanjing 210023, China}\\ 
       {jzhan9@uic.edu, szwang@cse.buaa.edu.cn, zhanqianyi@gmail.com, psyu@uic.edu}
}

%

\date{}
\maketitle

\begin{abstract}
Traditional viral marketing problems aim at selecting a subset of seed users for one single product to maximize its awareness in social networks. However, in real scenarios, multiple products can be promoted in social networks at the same time. At the product level, the relationships among these products can be quite intertwined, e.g., \textit{competing}, \textit{complementary} and \textit{independent}. In this paper, we will study the ``\textit{inter\underline{T}wined \underline{I}nfluence \underline{M}aximization}'' (i.e., {\problem}) problem for one product that we target on in online social networks, where multiple other competing/complementary/independent products are being promoted simultaneously. The {\problem} problem is very challenging to solve due to (1) few existing models can handle the \textit{intertwined diffusion} procedure of multiple products concurrently, and (2) optimal seed user selection for the target product may depend on other products' marketing strategies a lot. To address the {\problem} problem, a unified greedy framework {\our} (inter\underline{T}wined \underline{I}nfluence \underline{E}stimato\underline{R}) is proposed in this paper. Extensive experiments conducted on four different types of real-world social networks demonstrate that {\our} can outperform all the comparison methods with significant advantages in solving the {\problem} problem.
\end{abstract}




\section{Introduction}\label{sec:intro}

\begin{figure}[t]
\centering
    \begin{minipage}[l]{0.63\columnwidth}
      \centering
      \includegraphics[width=\textwidth]{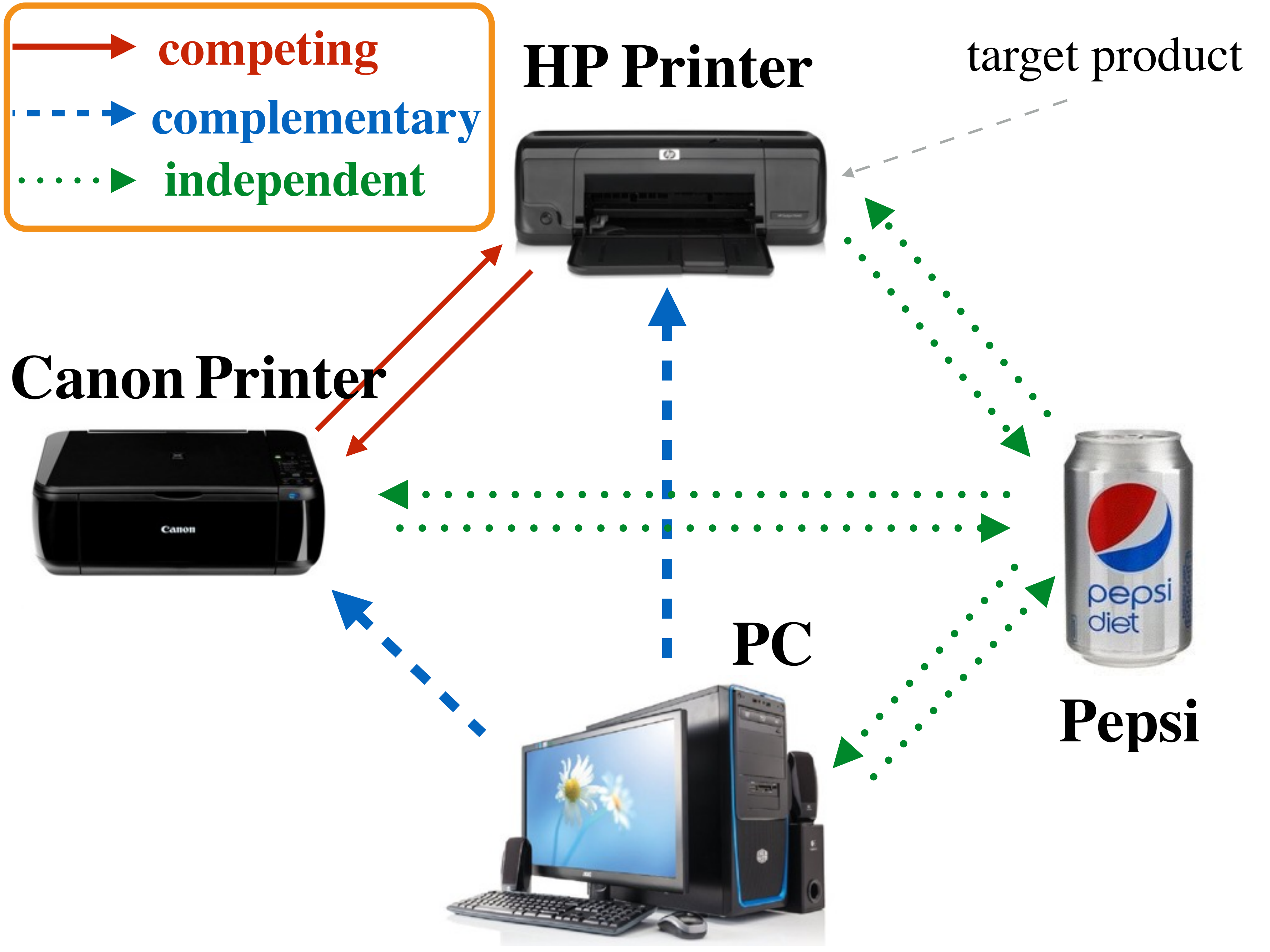}
    \end{minipage}
  \caption{Intertwined relationships among products.}\label{fig:relationships}
\end{figure}

Viral marketing (i.e., social influence maximization) first proposed in \cite{DR01} has become a hot research problem in recent years and dozens of papers on this topic have been published so far \cite{KKT03, LKGFVG07, CWY09, CWW10, SCHT07, GBL08}. Traditional viral marketing problem aims at selecting the optimal set of seed users to maximize the awareness of ideas or products in social networks and has extensive concrete applications in the real world, e.g., product promotion \cite{DMS10, NN12} and opinion spread \cite{CCCKLRSWWY11}. In the traditional viral marketing setting \cite{DR01, KKT03}, only one product/idea is to be promoted. However, in the real scenarios, the promotions of multiple products can co-exist in the social networks at the same time. For example, in Figure~\ref{fig:relationships}, we show $4$ different products to be promoted in an online social network and HP printer is our target product. At the product level, the relationships among these products can be quite intertwined:

\begin{itemize}

\item \textit{independent}: promotion activities of some products (e.g., HP printer and Pepsi) can be \textit{independent} of each other.

\item \textit{competing}: products having common functions will \textit{compete} for the market share \cite{BKS07, CNWZ07} (e.g., HP printer and Canon printer). Users who have bought a HP printer are less likely to buy a Canon printer again.

\item \textit{complementary}: product cross-sell is also very common in marketing \cite{NN12}. Users who have bought a certain product (e.g., PC) will be more likely to buy another product (e.g., HP printer) and the promotion of PC is said to be \textit{complementary} to that of HP printer.


\end{itemize}

\noindent \textbf{Problem}: In this paper, we want to maximize the influence of one specific product that we target on in online social networks, where many other products are being promoted simultaneously. The relationships among these product can be obtained in advance via effective market research, which can be \textit{independent}, \textit{competitive} or \textit{complementary}. Formally, we define this problem as the inter\underline{T}wined \underline{I}nfluence \underline{M}aximization ({\problem}) problem.



Before starting the promotions, companies need to design their \textit{marketing strategies} carefully. \textit{Marketing strategies} includes all basic and long-term activities in the field of marketing that can contribute to the goals of the company and its marketing objectives. However, in this paper, we are mainly concerned about the selected \textit{seed users} who will spread the influence in social networks. Hence, for simplicity, we refer to the \textit{marketing strategies} of products as the \textit{seed users} selected for the products. 

More specifically, depending on the promotional order of other products and the target product, the {\problem} problem can have two different variants (we don't care about the case that other products are promoted after the target product):

\begin{itemize}

\item \textit{{\problemknown} problem}: In some cases, the other products have been promoted ahead of the target products, where their selected seed users are known and product information has already been propagated within the network. In such a case, the variant of {\problem} is defined as the \underline{C}onditional inter\underline{T}wined \underline{I}nfluence \underline{M}aximization ({\problemknown}) problem.

\item \textit{{\problemunknown} problem}: However, in some other cases, the promotion activities of multiple products occur simultaneously, where the \textit{marketing strategies} of all these products are confidential to each other. Such a variant of {\problem} is defined as the \underline{J}oint inter\underline{T}wined \underline{I}nfluence \underline{M}aximization ({\problemunknown}) problem.

\end{itemize}

The {\problem} problem (both {\problemknown} and {\problemunknown}) studied in this paper is a novel problem and totally different from existing works on viral marketing: \textit{traditional single-product viral marketing problem} \cite{KKT03}, \textit{viral marketing for multiple independent products} \cite{DMS10}, \textit{viral marketing for competing products only} \cite{BKS07, CNWZ07, BFO10}, and \textit{viral marketing for cross-sell products only} \cite{NN12}. More information of other related problems is available in Section~\ref{sec:relatedwork}.

Despite its importance and novelty, the {\problem} problem is very challenging to solve due to the following reasons:

\begin{itemize}

\item \textit{Lack of information diffusion model}: A new diffusion model which can handle the intertwined diffusion of these \textit{independent}, \textit{competing} and \textit{complementary} products is the prerequisite for addressing the {\problem} problem.

\item \textit{Utilization of the known marketing strategies}: In the {\problemknown} problem, other products have been promoted in advance and their \textit{marketing strategies} are public already. How to utilize these known \textit{marketing strategies} to help identify the optimal seed user set for the target product is very challenging.

\item \textit{Unknown marketing strategies}: In the {\problemunknown} problem, \textit{marketing strategies} of other products are unknown. Inferring the potential \textit{marketing strategies} of these products and developing the optimal  \textit{marketing strategies} for the target product based on the inference is still an open problem to this context so far.


\end{itemize}

To solve all the above challenges, we propose a unified \textit{greedy} framework inter\underline{T}wined \underline{I}nfluence \underline{E}stimato\underline{R} ({\our}) in this paper. The {\our} method also has two variants: (1) {\ourknown} (\underline{C}onditional {\our}) for the {\problemknown} problem, and (2) {\ourunknown} (\underline{J}oint {\our}) for the {\problemunknown} problem. {\our} is based on a novel information diffusion model inter\underline{T}wined \underline{L}inear \underline{T}hreshold ({\diffusion}) introduced in this paper. {\diffusion} quantifies the \textit{impacts} among products with the \textit{intertwined threshold updating strategy} and can handle the intertwined diffusions of these products at the same time. To solve the {\problemknown} problem, {\ourknown} will select seed users greedily and is proved to achieve a $(1-\frac{1}{e})$-approximation to the optimal result. For the {\problemunknown} problem, we show that the theoretical influence upper and lower bounds calculation is \textit{NP-hard}. Alternatively, we formulate the {\problemunknown} problem as a game among different products and propose to infer the potential \textit{marketing strategies} of other products. The \textit{step-wise greedy} method {\ourunknown} can achieve promising results by selecting seed users wisely according to the inferred marketing strategies of other products.

The rest of this paper is organized as follows. In Section~\ref{sec:formulation}, we give the concept and problem definitions. In Section~\ref{sec:method}, the {\diffusion} diffusion model and {\our} method are introduced in details, which will be evaluated in Section~\ref{sec:experiment}. Finally, we give the related works in Section~\ref{sec:relatedwork} and conclude the paper in Section~\ref{sec:conclusion}.

\section{Problem Formulation} \label{sec:formulation}

In this section, we will define some important concepts and give the formulation of the {\problem} problem.

\subsection{Concept Definitions}

\noindent \textbf{Definition 1} (Social Network): An online \textit{social network} can be represented as $G = (\mathcal{V}, \mathcal{E})$, where $\mathcal{V}$ is the set of users and $\mathcal{E}$ contains the interactions among users in $\mathcal{V}$. The set of $n$ different products to be promoted in network $G$ can be represented as $\mathcal{P} = \{p^1, p^2, \cdots, p^n\}$.

\noindent \textbf{Definition 2} (User Status Vector): For a given product $p^j \in \mathcal{P}$, users who are influenced to buy $p^j$ are defined to be ``\textit{active}'' to $p^j$, while the remaining users who have not bought $p^j$ are defined to be ``\textit{inactive}'' to $p^j$. User $u_i$'s status towards all the products in $\mathcal{P}$ can be represented as ``\textit{user status vector}'' $\mb{s}_i = (s_i^1, s_i^2, \cdots, s_i^n)$, where $s_i^j$ is $u_i$'s status to product $p^j$. Users can be activated by multiple products at the same time (even competing products), i.e., multiple entries in \textit{status vector} $\mb{s}_i$ can be ``\textit{active}'' concurrently.

\noindent \textbf{Definition 3} (Independent, Competing and Complementary Products): Let $P(s_i^j = 1)$ (or $P(s^j_i)$ for simplicity) denote the probability that $u_i$ is activated by product $p^j$ and $P(s_i^j | s_i^k)$ be the conditional probability given that $u_i$ has been activated by $p^k$ already. For products $p^j, p^k \in \mathcal{P}$, the promotion of $p^k$ is defined to be (1) \textit{independent} to that of $p^j$ if $\forall u_i \in \mathcal{V}$, $P(s_i^j | s_i^k) = P(s_i^j)$, (2) \textit{competing} to that of $p^j$ if $\forall u_i \in \mathcal{V}$, $P(s_i^j | s_i^k) < P(s_i^j)$, and (3) \textit{complementary} to that of $p^j$ if $\forall u_i \in \mathcal{V}$, $P(s_i^j | s_i^k) > P(s_i^j)$.


\noindent \textbf{Definition 4} (Marketing Strategy): In this paper, we are mainly concerned about the \textit{seed user} selection problem. For simplicity, we refer to the \textit{marketing strategy} of product $p^j \in \mathcal{P}$ as the \textit{seed user set} $\mathcal{S}^j$ selected for $p^j$. And the \textit{marketing strategies} of all products in $\mathcal{P}$ can be represented as seed user set list $\mathcal{S} = (\mathcal{S}^1, \mathcal{S}^2, \cdots, \mathcal{S}^n)$.


\subsection{Problem Definition}

In traditional single-product viral marketing problems, the selected \textit{seed users} will propagate the influence of the target product in the network and the number of users get activated can be obtained with the \textit{influence function} $I: \mathcal{S} \to \mathbb{R}$, which maps the selected seed users to the number of influenced users.

Traditional one single product viral marketing problem aims at selecting the optimal seed users $\bar{\mathcal{S}}$ for the target product, who can achieve the maximum influence:
\begingroup\makeatletter\def\f@size{9}\check@mathfonts$$\bar{\mathcal{S}} = \arg_{\mathcal{S}} \max I(\mathcal{S}).$$\endgroup

However, in the {\problem} problem, promotions of multiple products in $\mathcal{P}$ co-exist simultaneously. The influence function of the target product $p^j \in \mathcal{P}$ depends on not only the seed user set $\mathcal{S}^j$ selected for itself but also the seed users of other products in $\mathcal{P} \setminus \{p^j\}$. Based on such a intuition, we formally define the \textit{conditional intertwined influence function}, \textit{joint intertwined influence function} and give the formulation of {\ourknown}, {\ourunknown} problems as follows.


\noindent \textbf{Definition 5} (Conditional Intertwined Influence Function): Let $\mathcal{S}^{-j} = (\mathcal{S}^1, \cdots, \mathcal{S}^{j-1}, \mathcal{S}^{j+1}, \cdots,  \mathcal{S}^{n})$ be the known seed user sets selected for all products in $\mathcal{P} \setminus \{p^j\}$, the \textit{influence function} of the target product $p^j$ given the known \textit{seed user sets} $\mathcal{S}^{-j}$ is defined as the \textit{conditional intertwined influence function}: $I(\mathcal{S}^j | \mathcal{S}^{-j}).$


\noindent \textbf{C-TIM Problem}: {\problemknown} problem aims at selecting the optimal \textit{marketing strategy} $\mathcal{\bar{S}}^j$ to maximize the \textit{conditional intertwined influence function} of $p^j$ in the network, i.e., 
\begingroup\makeatletter\def\f@size{9}\check@mathfonts$$\bar{\mathcal{S}^j} = \arg_{\mathcal{S}^j} \max I(\mathcal{S}^j | \mathcal{S}^{-j}).$$\endgroup

\noindent \textbf{Definition 6} (Joint Intertwined Influence Function): When the seed user sets of products $\mathcal{P} \setminus \{p^j\}$ are unknown, i.e., $\mathcal{S}^{-j}$ is not given, the \textit{influence function} of product $p^j$ together with other products in $\mathcal{P} \setminus \{p^j\}$ is defined as the \textit{joint intertwined influence function}: $I(\mathcal{S}^j; \mathcal{S}^{-j}).$

\noindent \textbf{J-TIM Problem}: {\problemunknown} problem aims at choosing the optimal \textit{marketing strategy} $\mathcal{\bar{S}}^j$ to maximize the \textit{joint intertwined influence function} of $p^j$ in the network, i.e., 
\begingroup\makeatletter\def\f@size{9}\check@mathfonts$$\bar{\mathcal{S}^j} = \arg_{\mathcal{S}^j} \max I(\mathcal{S}^j; \mathcal{S}^{-j}),$$\endgroup
where set $\mathcal{S}^{-j}$ can take any possible value.

\section{Proposed Method}
\label{sec:method}

In this section, we will introduce the {\our} framework in details: In Section~\ref{subsec:diffusion}, we propose a new diffusion model {\diffusion} to deal with the intertwined diffusion of multiple products. In Section~\ref{subsec:problem_method_known}, we analyze the {\problemknown} problem and show that the proposed greedy method {\ourknown} can achieve a $(1-\frac{1}{e})$-approximation of the optimal results. In Section~\ref{subsec:problem_method_unknown}, we study the {\problemunknown} problem and propose a new approach {\ourunknown} to address {\problemunknown} by formulating it as a game among multiple products.

\subsection{Intertwined Information Diffusion}\label{subsec:diffusion}


\subsubsection{Preliminary}

In traditional single-product \textit{linear threshold} (LT) model, user $u_i \in \mathcal{V}$ can influence his neighbor $u_k \in \Gamma(u_i)$ according to weight $w_{i,k} \ge 0$ ($w_{i,k} = 0$ if link $(u_k, u_i)$ doesn't exist or user $u_i$ is \textit{inactive}), where $\Gamma(u_i)$ represents the set of users following $u_i$ (i.e., users that $u_i$ can influence). Each user, e.g., $u_i$, is associated with a \textit{static threshold} $\theta_i$ uniformly chosen at random from interval $[0, 1]$, which represents the minimal required influence for $u_i$ to become \textit{active}. Initially, only users in the seed user set $\mathcal{S}$ are active and their influence will propagate within the network in discrete steps. At step $t$, all \textit{active} users in step $t-1$ remain \textit{active} and \textit{inactive} user, e.g., $u_i$, can be \textit{activated} if the influence received from the other users can exceed his threshold, i.e., \begingroup\makeatletter\def\f@size{8}\check@mathfonts$\sum_{u_l \in \Gamma_{out}(u_i)} w_{l,i} \ge \theta_i$\endgroup, where $\Gamma_{out}(u_i)$ represents the set of users that $u_i$ follows.

\subsubsection{Intertwined Linear Threshold Model (TLT)}

To depict the intertwined diffusions of multiple independent/competing/complementary products, we propose a new information diffusion model {\diffusion}. In the existence of multiple products $\mathcal{P}$, user $u_i$'s influence to his neighbor $u_k$ in promoting product $p^j$ can be represented as $w^j_{i,k} \ge 0$. Similar to the traditional LT model, in {\diffusion}, the influence of different products can propagate within the network step by step. User $u_i$'s threshold for product $p^j$ can be represented as $\theta^j$ and $u_i$ will be activated by his neighbors to buy product $p^j$ if 
\begingroup\makeatletter\def\f@size{8}\check@mathfonts$$\sum_{u_l \in \Gamma_{out}(u_i)} w^j_{l,i} \ge \theta^j_i.$$\endgroup

Different from traditional LT model, in {\diffusion}, users in online social networks can be activated by multiple products at the same time, which can be either \textit{independent}, \textit{competing} or \textit{complementary}. As shown in Figure~\ref{fig:relationships}, we observe that users' chance to buy the HP printer will be (1) unchanged given that they have bought Pepsi (i.e., the \textit{independent} product of HP printer), (2) increased if they own PCs (i.e., the \textit{complementary} product of HP printer), and (3) decreased if they already have the Canon printer (i.e., the \textit{competing} product of HP printer).

To model such a phenomenon in {\diffusion}, we introduce the following \textit{intertwined threshold updating strategy}, where users' \textit{thresholds} to different products will change \textit{dynamically} as the influence of other products propagates in the network.


\noindent \textbf{Definition 7} (Intertwined Threshold Updating Strategy): Assuming that user $u_i$ has been activated by $m$ products $p^{\tau_1}$, $p^{\tau_2}$, $\cdots$, $p^{\tau_m} \in \mathcal{P} \setminus \{p^j\}$ in a sequence, then $u_i$'s \textit{threshold} towards product $p^j$ will be updated as follows:
\begingroup\makeatletter\def\f@size{8}\check@mathfonts \begin{align*}
(\theta^j_i)^{\tau_1} &= \theta^j_i \frac{P(s^j_i)}{P(s^j_i | s^{\tau_1}_i)}, (\theta^j_i)^{\tau_2} = (\theta^j_i)^{\tau_1} \frac{P(s^j_i | s^{\tau_1}_i)}{P(s^j_i | s^{\tau_1}_i, s^{\tau_2}_i)}, \cdots\\
(\theta^j_i)^{\tau_m} &= (\theta^j_i)^{\tau_{m-1}} \frac{P(s^j_i | s^{\tau_1}_i, \cdots, s^{\tau_{m-1}}_i)}{P(s^j_i | s^{\tau_1}_i, \cdots, s^{\tau_{m-1}}_i, s^{\tau_{m}}_i)},
\end{align*}\endgroup
where $(\theta^j_i)^{\tau_k}$ denotes $u_i$'s threshold to $p^j$ after he has been activated by $p^{\tau_1}$, $p^{\tau_2}$, $\cdots$, $p^{\tau_k}$, $k \in \{1, 2, \cdots, m\}$.


In this paper, we do not focus on the order of products that activate users \cite{CCCKLRSWWY11} and to simplify the calculation of the \textit{threshold updating strategy}, we assume only the most recent activation has an effect on updating current thresholds, i.e., 
\begingroup\makeatletter\def\f@size{8}\check@mathfonts$$\frac{P(s^j_i | s^{\tau_1}_i, \cdots, s^{\tau_{m-1}}_i)}{P(s^j_i | s^{\tau_1}_i, \cdots, s^{\tau_{m-1}}_i, s^{\tau_{m}}_i)} \approx \frac{P(s^j_i)}{P(s^j_i | s^{\tau_{m}}_i)}=\phi_i^{\tau_{m} \to j}.$$\endgroup

\noindent \textbf{Definition 8} (Threshold Updating Coefficient): Term $\phi_i^{l \to j} = \frac{P(s^j_i)}{P(s^j_i | s^l_i)}$ is formally defined as the ``\textit{threshold updating coefficient}'' of product $p^l$ to product $p^j$ for user $u_i$, where
\begingroup\makeatletter\def\f@size{8}\check@mathfonts$$
\phi_i^{l \to j}
\begin{cases} 
< 1,  & \mbox{if }p^l\mbox{ is \textit{complementary} to }p^j, \\
= 1,  & \mbox{if }p^l\mbox{ is \textit{independent} to }p^j, \\
> 1,  & \mbox{if }p^l\mbox{ is \textit{competing} to }p^j.
\end{cases}
$$\endgroup

The \textit{intertwined threshold updating strategy} can be rewritten based on the \textit{threshold updating coefficients} as follows:
\begingroup\makeatletter\def\f@size{8}\check@mathfonts$$(\theta^j_i)^{\tau_m} \approx \theta^j_i \cdot \phi_i^{\tau_{1} \to j} \cdot \phi_i^{\tau_{2} \to j} \cdots \phi_i^{\tau_{m} \to j}.$$\endgroup

\subsection{Conditional Intertwined Influence Maximization}\label{subsec:problem_method_known}


In the {\problemknown} problem, the promotion activities of other products have been done before we start to promote our target product. Subject to the {\diffusion} diffusion model, users' thresholds to the target product can be updated with the \textit{threshold updating strategy} after the promotions of other products. Based on the updated network, the {\problemknown} can be mapped to the \textit{tradition single-product viral marketing}, which has been proved to be \textit{NP-hard} already.

\begin{theorem}
{
The {\problemknown} problem is \textit{NP-hard} based on the {\diffusion} diffusion model.
}
\end{theorem}
The proof of Theorem 1 is omitted due to limited space.

Meanwhile, based on the {\diffusion} diffusion model, the \textit{conditional influence function} of the target product $I(\mathcal{S}^j | \mathcal{S}^{-j})$ are observed to be both \textit{monotone} and \textit{submodular}.

\begin{theorem}
{
For the {\diffusion} diffusion model, the \textit{conditional influence function} is \textit{monotone}.
}
\end{theorem}
\noindent \textit{Proof}: Given the existing seed user sets $\mathcal{S}^{-j}$ for existing products $\mathcal{P} - \{p^j\}$ in the market, let $\mathcal{T}$ be a seed user set of product $p^j$. Users in the network who are not involved in $\mathcal{T}$ can be represented as $\mathcal{V} - \mathcal{T}$. For the given seed user set $\mathcal{T}$ and the fixed seed users set $\mathcal{S}^{-j}$ of other products, adding a new seed user, e.g., $u \in \mathcal{V} - \mathcal{T}$, to the seed user set $\mathcal{T}$ will not decrease the number of influenced users, i.e., $I(\mathcal{T} \cup \{u\} | \mathcal{S}^{-j}) \ge I(\mathcal{T} | \mathcal{S}^{-j})$.

\begin{theorem}
{
For the {\diffusion} diffusion model, the \textit{conditional influence function} is \textit{submodular}.
}
\end{theorem}
\noindent \textit{Proof}: After the diffusion process of the existing products in $\mathcal{P}-\{p^j\}$, users the thresholds towards product $p^j$ will be updated. Based on the updated network, for two given seed user sets $\mathcal{R}$ and $\mathcal{T}$, where $\mathcal{R} \subseteq \mathcal{T} \subseteq \mathcal{V}$, it is easy to show that $I(\mathcal{R} \cup \{v\} | \mathcal{S}^{-j}) - I(\mathcal{R} |\mathcal{S}^{-j}) \ge I(\mathcal{T} \cup \{v\} | \mathcal{S}^{-j}) - I(\mathcal{T} | \mathcal{S}^{-j})$ with the ``\textit{live-edge path}'' \cite{KKT03}.



\begin{algorithm}[t]
\scriptsize
\caption{The {\ourknown} Algorithm}
\label{alg:s}
\begin{algorithmic}[1]
	\REQUIRE input social network $G=(\mathcal{V}, \mathcal{P}, \mathcal{E})$\\
\qquad target product: $p^j$\\
\qquad known seed user sets of $\mathcal{P} - \{p^j\}$: $\mathcal{S}^{-j}$\\
\qquad conditional influence function of $p^j$: $I(\mathcal{S}^j | \mathcal{S}^{-j})$\\
\qquad seed user set size of $p^j$: $k^j$
\ENSURE  selected seed user set $\mathcal{S}^j$ of size $k^j$


\STATE	{initialize seed user set $\mathcal{S}^j = \emptyset$}\\
\STATE	{propagate influence of products $\mathcal{P} - \{p^j\}$ with $\mathcal{S}^{-j}$ and update users' thresholds with intertwined threshold updating strategy}\\
\WHILE	{$\mathcal{V} \setminus \mathcal{S}^j \neq \emptyset \land \left | \mathcal{S}^j \right | \neq k^j$}

\STATE	{pick a user $u \in \mathcal{V} - \mathcal{S}^j$ according to equation $\arg \max_{u \in \mathcal{V}} I(\mathcal{S}^j \cup \{u\} | \mathcal{S}^{-j}) - I(\mathcal{S}^j | \mathcal{S}^{-j})$}
\STATE	{$\mathcal{S}^j = \mathcal{S}^j \cup \{u\}$}
\ENDWHILE
\STATE	{return $\mathcal{S}^j$.}

\end{algorithmic}
\end{algorithm}

According to the above analysis, a greedy algorithm {\ourknown} is proposed to solve the problem {\problemknown} in this paper, whose pseudo code is available in Algorithm~\ref{alg:s}. In {\ourknown}, we select the user $u$ who can lead to the maximum increase of the conditional influence function $I(\mathcal{S}^j \cup \{u\} | \mathcal{S}^{-j})$ at each step as the new seed user. This process repeats until either no potential seed user is available or all the $k^j$ required seed users have been selected. The time complexity of {\ourknown} is $O(k^j |\mathcal{V}| (|\mathcal{V}| + |\mathcal{E}|))$. Since the \textit{conditional influence function} is \textit{monotone} and \textit{submodular} based on the {\diffusion} diffusion model, then the \textit{step-wise greedy} algorithms {\ourknown}, which select the users who can lead to the maximum increase of influence, can achieve a $(1 - \frac{1}{e})$-approximation of the optimal result for the target product.


\subsection{Joint Intertwined Influence Maximization}\label{subsec:problem_method_unknown}

{\problemknown} studies a common case in real-world viral marketing, where different companies have different schedules to release the promote their products and some can be conducted ahead of the target product. Meanwhile, in this section, we will study a more challenging case: {\problemunknown}, where other products are being promoted at the same time as our target product and the marketing strategies of different products are totally confidential.


\subsubsection{The {\problemunknown} Problem}

When the \textit{marketing strategies} of other products are unknown, the \textit{influence function} of the target product and other products co-exist in the network is defined as the \textit{joint influence function}: $I(\mathcal{S}^j; \mathcal{S}^{-j})$. Meanwhile, by setting $\mathcal{S}^1 = \cdots = \mathcal{S}^{j-1} = \mathcal{S}^{j+1} = \cdots = \mathcal{S}^n = \emptyset$, the {\problemunknown} problem can be mapped to the traditional \textit{single-product influence maximization} problem in polynomial time, which is an NP-hard problem.

\begin{theorem}
{
The {\problemunknown} problem is \textit{NP-hard} based on the {\diffusion} diffusion model.
}
\end{theorem}\label{theo:jtimnp}



Meanwhile, if all the products in $\mathcal{P} \setminus \{p^j\}$ are \textit{independent} to $p^j$, the \textit{joint influence function} $I(\mathcal{S}^j; \mathcal{S}^{-j})$ will be both \textit{monotone} and \textit{submodular}.

\begin{theorem}\label{th:monotone}
{
Based on the {\diffusion} diffusion model, the \textit{joint influence function} is \textit{monotone} and \textit{submodular} if all the other products are \textit{independent} to $p^j$.
}
\end{theorem}

However, when there exist products in $\mathcal{P} \setminus \{p^j\}$ to be either \textit{competing} or \textit{complementary} to $p^j$, the \textit{joint influence function} $I(\mathcal{S}^j; \mathcal{S}^{-j})$ will be neither \textit{monotone} nor \textit{submodular}.

\begin{theorem}\label{th:not_monotone}
{
Based on the {\diffusion} diffusion model, the \textit{joint influence function} is not \textit{monotone} if there exist products which are either \textit{competing} or \textit{complementary} to the target product $p^j$.
}
\end{theorem}

\begin{theorem}\label{th:not_submodular}
{
For the {\diffusion} diffusion model, the \textit{joint influence function} is not \textit{submodular} if these exist products which are either \textit{competing} or \textit{complementary} to to the target product $p^j$.
}
\end{theorem}

\noindent The proof of Theorem 4 is omitted due to the limited space and those of Theorems 5-7 are available in Appendix A-C.

\subsubsection{Challenges in {\problemunknown}}

\begin{algorithm}[t]
\scriptsize
\caption{The {\ourunknown} Algorithm}
\label{alg:m}
\begin{algorithmic}[1]

\REQUIRE input social network $G=(\mathcal{V}, \mathcal{P}, \mathcal{E})$\\
\qquad target product: $p^j$\\
\qquad set of other products: $\mathcal{P} - \{p^j\}$\\
\qquad joint influence function of $p^j$: $I(\mathcal{S}^j; \mathcal{S}^{-j})$\\
\qquad seed user set size of products in $\mathcal{P}$:$k^1, k^2, \cdots, k^j, \cdots, k^n$
\ENSURE  selected seed user sets $\{{\mathcal{S}}^1, {\mathcal{S}}^2, \cdots, {\mathcal{S}}^n\}$ of products in $\mathcal{P}$ respectively


\STATE	{initialize seed user set ${\mathcal{S}}^1, \mathcal{S}^2, \cdots, {\mathcal{S}}^n = \emptyset$}\\
\WHILE	{$(\mathcal{V} \setminus \mathcal{S}^1 \neq \emptyset \lor \cdots \lor \mathcal{V} \setminus \mathcal{S}^n \neq \emptyset) \land (\left | \mathcal{S}^1 \right | \neq k^1 \lor \cdots \lor \left | \mathcal{S}^n \right | \neq k^n)$}
\FOR		{random $i \in \{1, 2, \cdots, n\}$ ($p^i$ has not selected seeds in the round yet)}
\IF		{$\mathcal{V} \setminus \mathcal{S}^i \neq \emptyset \land \left | \mathcal{S}^i \right | \neq k^i$}
\STATE	{$p^i$ infers the seed user sets $\mathcal{\bar{S}}^{-i}$ of other products}
\STATE	{$p^i$ selects its seed user $u^i \in \mathcal{V} - {\mathcal{S}}^i$, who can maximize $I(\mathcal{S}^i \cup \{u^i\}; \mathcal{\bar{S}}^{-i}) - I(\mathcal{S}^i; \mathcal{\bar{S}}^{-i})$}
\STATE	{$\mathcal{S}^i = \mathcal{S}^i \cup \{u^i\}$}
\STATE	{propagate influence of $u$ in $G$ and update influenced users' thresholds to products in $\mathcal{P}$ with the \textit{intertwined threshold updating strategy}.}
\ENDIF
\ENDFOR
\ENDWHILE
\STATE	{return $\mathcal{S}^1, \mathcal{S}^2, \cdots, \mathcal{S}^n$.}

\end{algorithmic}
\end{algorithm}

When all the other products are \textit{independent} to $p^j$, the \textit{joint influence function} of $p^j$ will be \textit{monotone} and \textit{submodular}, which is solvable with the \textit{traditional greedy algorithm} proposed \cite{KKT03} and can achieve $(1-\frac{1}{e})$-approximation of the optimal results. However, when there exist at least one product which is either \textit{competing} or \textit{complementary} to $p^j$, the \textit{joint influence function} will be no longer \textit{monotone} or \textit{submodular}. In such a case, the {\problemunknown} will be very hard to solve and no promising optimality bounds of the results are available.

By borrowing ideas from the game theory studies \cite{NRTV07, BCIKKMP10}, for product $p^j$, the lower-bound and upper-bound of influence the {\problemunknown} problem can be achieved by selecting seed users of size $k$ can be represented as
\begingroup\makeatletter\def\f@size{8}\check@mathfonts$$\max_{\mathcal{S}^j} \min_{\mathcal{S}^{-j}} I(\mathcal{S}^j; \mathcal{S}^{-j}),\ \ \max_{\mathcal{S}^j} \max_{\mathcal{S}^{-j}} I(\mathcal{S}^j; \mathcal{S}^{-j})$$\endgroup
respectively, which denotes the maximum influence $p^j$ can achieve in the worst (and the best) cases where all the remaining products work together to make $p^j$'s influence as low (and high) as possible. The \textit{seed user set} selected by $p^j$ when achieving the lower-bound and upper-bound of influence can be represented as
\begingroup\makeatletter\def\f@size{8}\check@mathfonts$$\hat{\mathcal{S}}^j_{low} = \arg \max_{\mathcal{S}^j} \min_{\mathcal{S}^{-j}} I(\mathcal{S}^j; \mathcal{S}^{-j}),\ \hat{\mathcal{S}}^j_{up} = \arg \max_{\mathcal{S}^j} \max_{\mathcal{S}^{-j}} I(\mathcal{S}^j; \mathcal{S}^{-j}).$$\endgroup

However, the lower and upper bounds of the optimal results of the {\problemunknown} problem is hard to calculate mathematically.

\begin{theorem}
{
Computing the Max-Min for 3 or more player games is NP-hard.
}
\end{theorem}

\noindent \textit{Proof}: As proposed in \cite{BCIKKMP10}, the problem of finding any (approximate) Nash equilibrium for a three-player game is computationally intractable and it is NP-hard to approximate the min-max payoff value for each of the player \cite{BCIKKMP10, DGP06, CDT06, CTV07}.

\subsubsection{The {\ourunknown} Algorithm}

In addition, in the real world, the other products will not co-operate together in designing their marketing strategies to create the worst or the best situations for the target product $p^j$, i.e., choosing the \textit{marketing strategies}
 $\mathcal{S}^{-j}$ such that the \textit{joint influence function} $I(\mathcal{S}^j; \mathcal{S}^{-j})$ is minimized or maximized. To address the {\problemunknown} problem, in this part, we propose the {\ourunknown} algorithm to simulate the intertwined round-wise greedy seed user selection process of all the products.

In {\ourunknown}, all products are assumed to be \textit{selfish} and wants to maximize their own influence when selecting seed users based on the ``\textit{current}'' situation created by all the products. {\ourunknown} will infer the next potential \textit{marketing strategies} of other products round by round and select the \textit{optimal} seed users for each product based on the inference.

In algorithm {\ourunknown}, we let all products in $\mathcal{P}$ choose their optimal \textit{seed users} randomly at each round. For example, let $(\mathcal{S})^{\tau -1}$ be the seed users selected by products in $\mathcal{P}$ at round $\tau - 1$. At round $\tau$, a random product $p^i$ can select one seed user. To achieve the largest influence, product $p^i$ will infer the next potential seed users to be selected by other products based on the assumption that they are all selfish. For example, based $p^i$'s inference, the next seed user to be selected by $p^j$ can be represented as $\bar{u}^j$, i.e.,
\begingroup\makeatletter\def\f@size{6}\check@mathfonts \begin{align*}
\arg \max_{u \in \mathcal{V} - (\mathcal{S}^j)^{\tau - 1}} [ I \left ((\mathcal{S}^j)^{\tau - 1} \cup \{u\}; (\mathcal{S}^{-j})^{\tau - 1} \right) - I \left ( (\mathcal{S}^j)^{\tau - 1}; (\mathcal{S}^{-j})^{\tau - 1} \right)].
\end{align*}\endgroup

Similarly, $p^i$ can further infer the potential seed users to be selected next by products in $\mathcal{P}\setminus \{p^i, p^j\}$, who can be represented as $\{\bar{u}_1, \bar{u}_2, \cdots, \bar{u}_{i-1}, \bar{u}_{i+1}, \cdots, \bar{u}_{j-1}, \bar{u}_{j+1}, \cdots, \bar{u}_{n}\}$ respectively. Based on such inference, $p^i$ knows who are the next seed users to be selected by other products and will make use of the ``prior knowledge'' to select its own seed user $\hat{u}^i$ in round $\tau$:
\begingroup\makeatletter\def\f@size{7}\check@mathfonts \begin{align*}
\hat{u}^i = \arg \max_{u \in \mathcal{V} - (\mathcal{S}^i)^{\tau - 1}} [ I \left ((\mathcal{S}^i)^{\tau - 1} \cup \{u\}; \mathcal{\bar{S}}^{-i} \right) - I \left ( (\mathcal{S}^i)^{\tau - 1}; \mathcal{\bar{S}}^{-i} \right )].
\end{align*}\endgroup
where $\mathcal{\bar{S}}^{-i}$ is the ``inferred'' seed user sets of other products inferred by $p^i$ based on current situation by ``adding'' these inferred potential seed users to their seed user sets.

The selected $(\hat{u}^i)^{\tau}$ will be added to the seed user set of product $p^i$, i.e.,
\begingroup\makeatletter\def\f@size{8}\check@mathfonts$$(\mathcal{S}^i)^{\tau} = (\mathcal{S}^i)^{\tau - 1} \cup \{(\hat{u}^i)^{\tau}\}.$$\endgroup
And the ``\textit{current}'' seed user sets of all the products, i.e., $\mathcal{S}$, is updated as follows:
\begingroup\makeatletter\def\f@size{8}\check@mathfonts$$\mathcal{S} = ((\mathcal{S}^1)^{\tau}, (\mathcal{S}^2)^{\tau - 1}, \cdots, (\mathcal{S}^n)^{\tau - 1}).$$\endgroup

The selected $(\hat{u}^i)^{\tau}$ will propagate his influence in the network and all the users just activated to product $p^i$ will update their thresholds to other products in $\mathcal{P} \setminus \{p^i\}$.

Next, we let another random product (which has not selected seed users yet) to infer the next seed users to be selected by other products and choose its seed user based on the inferred situation. In each round, each product will have a chance to select one seed user and the user selection order of different products in each round is totally random. Such a process will stop when all the products either have selected the required number of \textit{seed users} or no users are available to be chosen. With the {\ourunknown} model, we simulate an alternative seed user selection procedure of multiple products in viral marketing and the pseudo-code {\ourunknown} method is given in Algorithm~\ref{alg:m}. The time complexity of the {\ourunknown} algorithm is $O((\sum_i k_i \cdot n)|\mathcal{V}|(|\mathcal{V}| + |\mathcal{E}|)$, where $k_i = |\mathcal{S}^i|$ is the number of seed users to be selected for product $p^i$.

\section{Experiments}\label{sec:experiment}

\begin{table}[t]
\scriptsize
\caption{Properties of the Different Networks}
\label{tab:datastat}
\centering
\begin{tabular}{clrr}
\toprule
\textbf{network} &\textbf{\# nodes} &\textbf{\# links} &\textbf{link type}\\
\midrule
Facebook &4,039 & 88,234 & undirected\\
\midrule
Wikipedia &7,115 &103,689 & directed\\
\midrule
arXiv &5,242 &14,496 & undirected\\
\midrule
Epinions &7,725 &82,861 & directed\\
\bottomrule
\end{tabular}
\end{table}

\begin{figure*}[t]
\centering
\subfigure[Facebook]{\label{eg_fig11_1}
    \begin{minipage}[l]{0.47\columnwidth}
      \centering
      \includegraphics[width=1.0\textwidth]{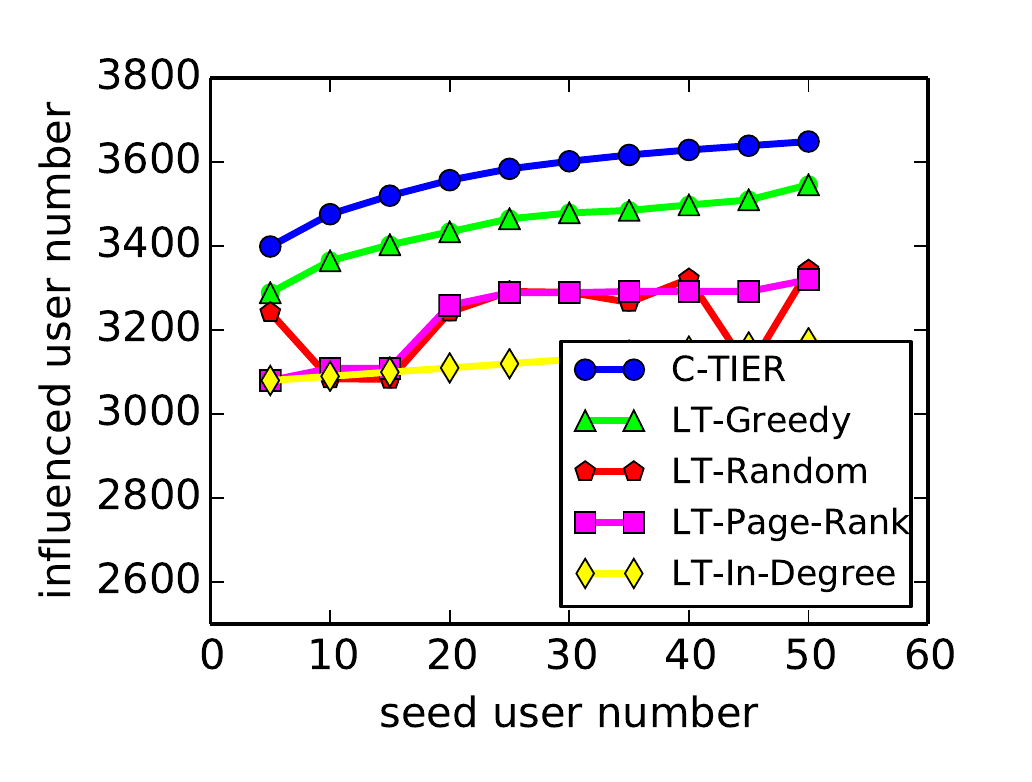}
    \end{minipage}
}
\subfigure[Wikipedia]{\label{eg_fig11_2}
    \begin{minipage}[l]{0.47\columnwidth}
      \centering
      \includegraphics[width=1.0\textwidth]{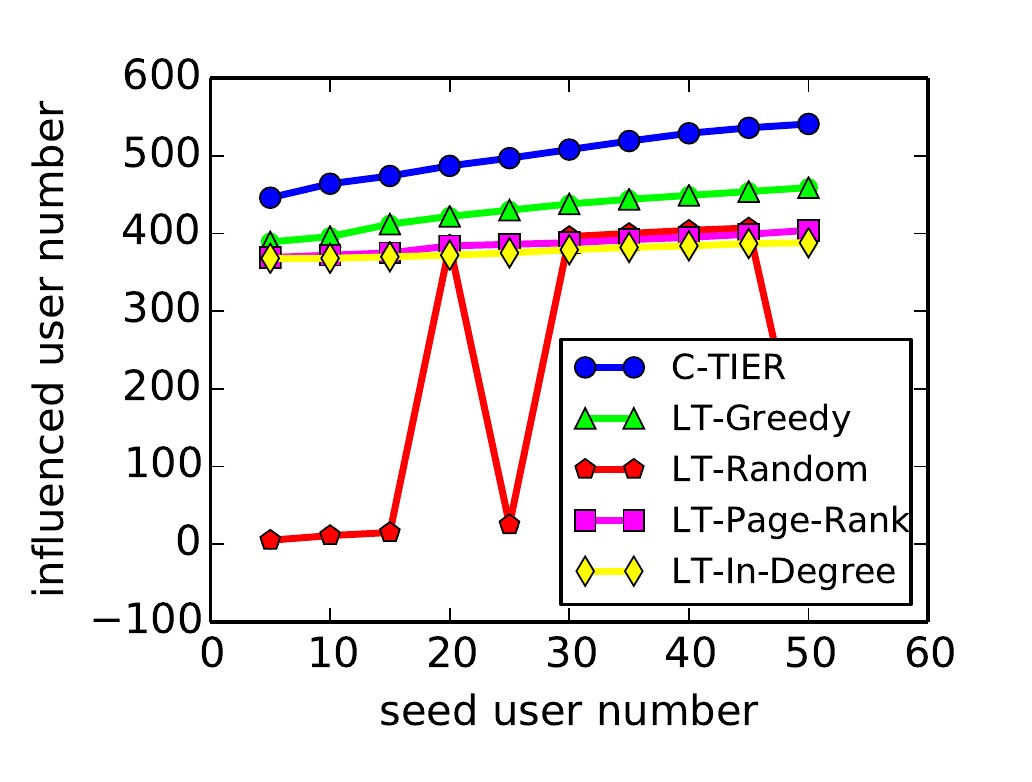}
    \end{minipage}
}
\subfigure[arXiv]{\label{eg_fig11_3}
    \begin{minipage}[l]{0.47\columnwidth}
      \centering
      \includegraphics[width=1.0\textwidth]{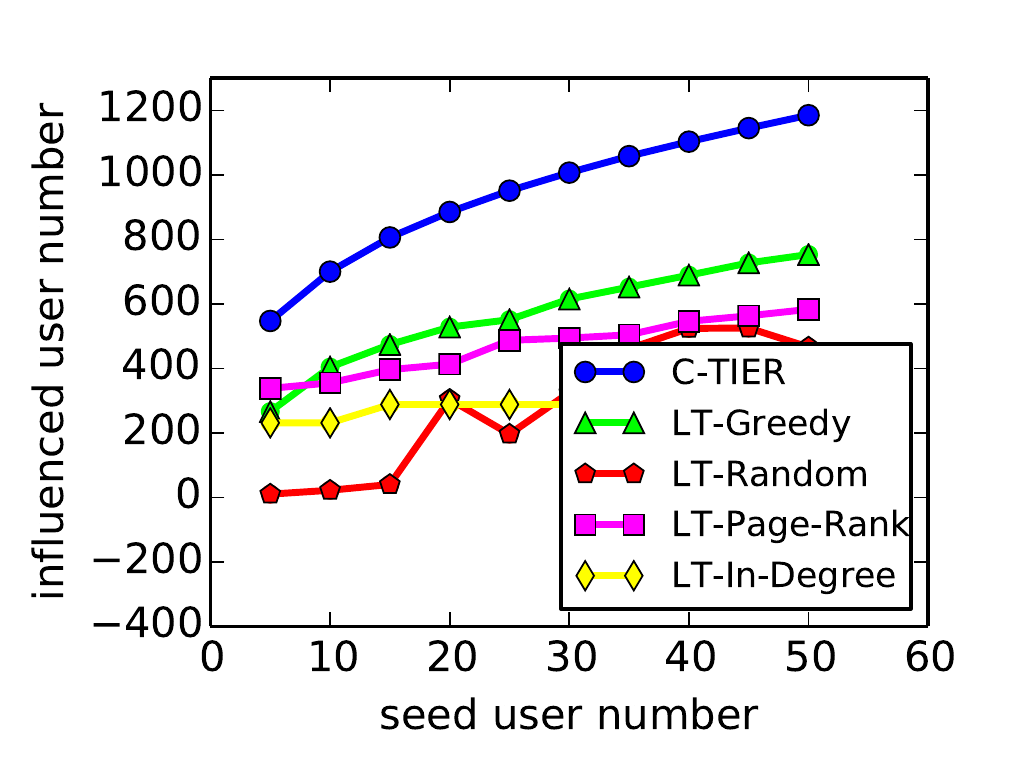}
    \end{minipage}
}
\subfigure[Epinions]{\label{eg_fig11_4}
    \begin{minipage}[l]{0.47\columnwidth}
      \centering
      \includegraphics[width=1.0\textwidth]{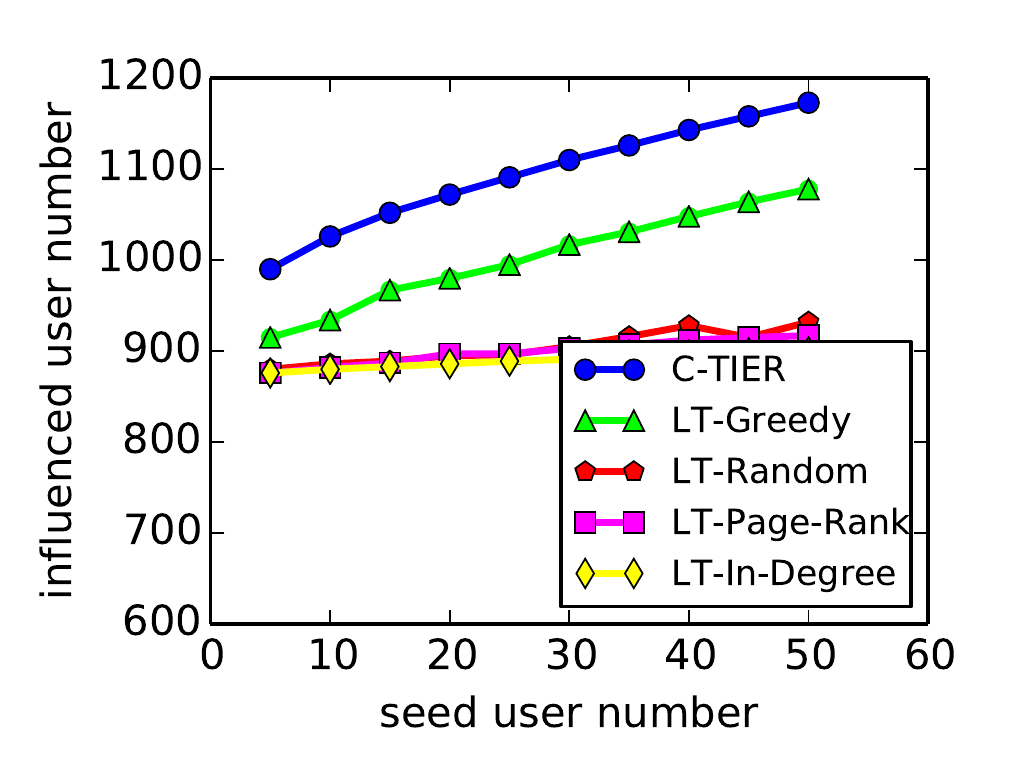}
    \end{minipage}
}
\caption{Experiment results of the {\problemknown} problem.}\label{eg_fig11}
\end{figure*}


\begin{table*}[t]
\scriptsize
\centering
{
\caption{Intersection of seed users selected by different comparison methods in the {\problemknown} problem.}\label{tab:intersection}

\begin{tabular}{ccccccccccc}

{\ourknown} & LT-G & LT-P & LT-I & LT-R & & LT-R & LT-I & LT-P & LT-G &{\ourknown}\\

\cline{1-5} \cline{7-11}
\multicolumn{1}{ |c }{50} & \multicolumn{1}{ |c }{31} &\multicolumn{1}{ |c }{0} &\multicolumn{1}{ |c }{0} &\multicolumn{1}{ |c| }{0} &{\ourknown} &\multicolumn{1}{ |c }{1} &\multicolumn{1}{ |c }{0} &\multicolumn{1}{ |c }{1} &\multicolumn{1}{ |c }{26} &\multicolumn{1}{ |c| }{50}\\

\cline{1-5} \cline{7-11}

& \multicolumn{1}{ |c }{50} &\multicolumn{1}{ |c }{1} &\multicolumn{1}{ |c }{0} &\multicolumn{1}{ |c| }{0} &LT-G &\multicolumn{1}{ |c }{1} &\multicolumn{1}{ |c }{0} &\multicolumn{1}{ |c }{0} &\multicolumn{1}{ |c| }{50} & \\

\cline{2-5} \cline{7-10}

& &\multicolumn{1}{ |c }{50} &\multicolumn{1}{ |c }{5} &\multicolumn{1}{ |c| }{0} &LT-P &\multicolumn{1}{ |c }{0} &\multicolumn{1}{ |c }{32} &\multicolumn{1}{ |c| }{50} & & \\

\cline{3-5} \cline{7-9}

& & &\multicolumn{1}{ |c }{50} &\multicolumn{1}{ |c| }{1} &LT-I &\multicolumn{1}{ |c }{0} &\multicolumn{1}{ |c| }{50} & & & \\

\cline{4-5} \cline{7-8}

\multicolumn{3}{ c }{Facebook} & &\multicolumn{1}{ |c| }{50} &LT-R  &\multicolumn{1}{ |c| }{50} & & \multicolumn{3}{ c }{Wikipedia} \\

\cline{5-5} \cline{7-7}

\\

{\ourknown} & LT-G & LT-P & LT-I & LT-R & & LT-R & LT-I & LT-P & LT-G &{\ourknown}\\

\cline{1-5} \cline{7-11}
\multicolumn{1}{ |c }{50} & \multicolumn{1}{ |c }{23} &\multicolumn{1}{ |c }{2} &\multicolumn{1}{ |c }{0} &\multicolumn{1}{ |c| }{1} &{\ourknown} &\multicolumn{1}{ |c }{0} &\multicolumn{1}{ |c }{0} &\multicolumn{1}{ |c }{0} &\multicolumn{1}{ |c }{30} &\multicolumn{1}{ |c| }{50}\\

\cline{1-5} \cline{7-11}

& \multicolumn{1}{ |c }{50} &\multicolumn{1}{ |c }{2} &\multicolumn{1}{ |c }{0} &\multicolumn{1}{ |c| }{0} &LT-G &\multicolumn{1}{ |c }{1} &\multicolumn{1}{ |c }{0} &\multicolumn{1}{ |c }{0} &\multicolumn{1}{ |c| }{50} & \\

\cline{2-5} \cline{7-10}

& &\multicolumn{1}{ |c }{50} &\multicolumn{1}{ |c }{5} &\multicolumn{1}{ |c| }{1} &LT-P &\multicolumn{1}{ |c }{1} &\multicolumn{1}{ |c }{36} &\multicolumn{1}{ |c| }{50} & & \\

\cline{3-5} \cline{7-9}

& & &\multicolumn{1}{ |c }{50} &\multicolumn{1}{ |c| }{1} &LT-I &\multicolumn{1}{ |c }{1} &\multicolumn{1}{ |c| }{50} & & & \\

\cline{4-5} \cline{7-8}

\multicolumn{3}{ c }{arXiv} & &\multicolumn{1}{ |c| }{50} &LT-R  &\multicolumn{1}{ |c| }{50} & & \multicolumn{3}{ c }{Epinions} \\

\cline{5-5} \cline{7-7}

\end{tabular}
}
\end{table*}

%


Considering that real-world social networks with multiple competing, complementary and independent products being promoted simultaneously is extremely difficult to obtain. To test the effectiveness of {\our} in addressing the {\problem} problem, we will conduct extensive experiments on $4$ real-world social network datasets, where $4$ generated products with intertwined relationships will be promoted simultaneously. This section contains $5$ parts: (1) dataset descriptions, (2) experiment setting of the {\problemknown} problem, (3) experiment results of the {\problemknown} problem, (4) experiment setting of the {\problemunknown} problem, and (5) experiment results of the {\problemunknown} problem.

\subsection{Dataset Description}

The datasets used in the experiment include (1) Facebook social network\footnote{http://snap.stanford.edu/data/egonets-Facebook.html}, (2) Wikipedia administrator vote network\footnote{http://snap.stanford.edu/data/wiki-Vote.html}, (3) arXiv collaboration network\footnote{http://snap.stanford.edu/data/ca-GrQc.html}, and (4) Epinions e-commerce trust network\footnote{http://www.public.asu.edu/~jtang20/datasetcode/truststudy.htm}. These $4$ different network datasets are all public and of different categories, which include the widely used social networks Facebook (where various social influence can diffuse among users), vote network (where voters' opinions about candidates could diffuse), academic co-author network (where academic ideas can propagate among researchers), and e-commerce network (where customers' reviews of products can influence other customers). Some statistical information about these $4$ datasets is given in Tables~\ref{tab:datastat}. More detailed information about these datasets is available at their corresponding webpages.

\noindent \textbf{Repeatability}: All these datasets are public and can be downloaded. The code of the experiments is available at link\footnote{https://www.dropbox.com/s/imi7625awnezrbx/expHybridIM.tar.gz?dl=0}.

\begin{figure*}[t]
\centering
\subfigure[Facebook]{\label{eg_fig12_1}
    \begin{minipage}[l]{0.47\columnwidth}
      \centering
      \includegraphics[width=1.0\textwidth]{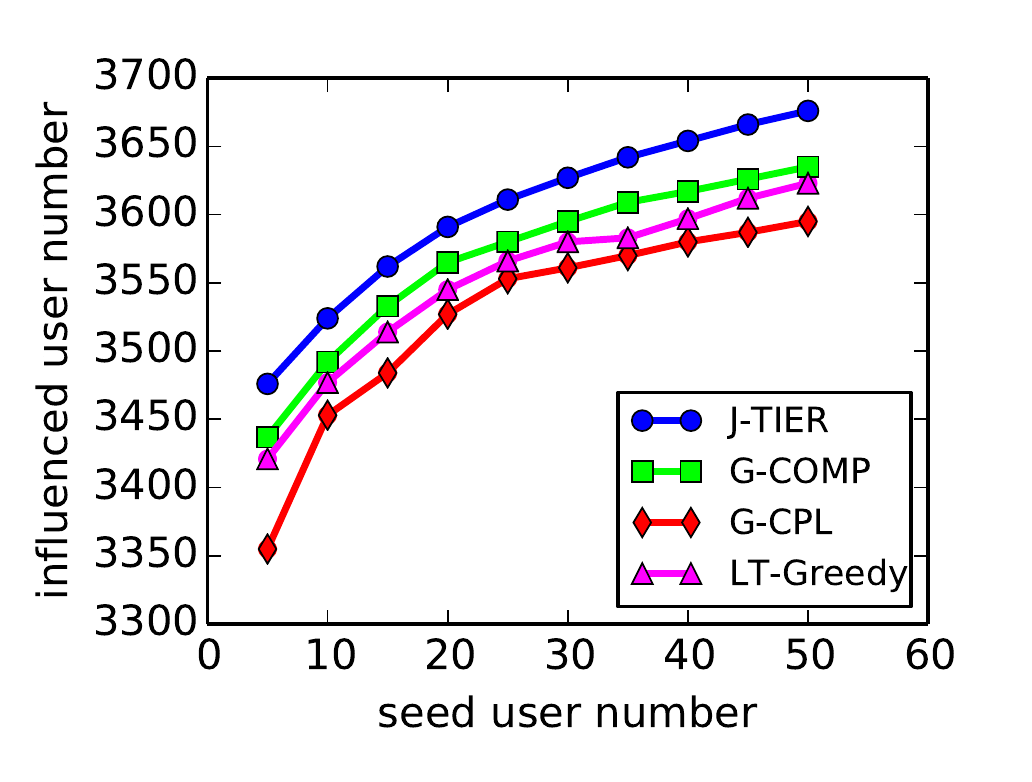}
    \end{minipage}
}
\subfigure[Wikipedia]{\label{eg_fig12_2}
    \begin{minipage}[l]{0.47\columnwidth}
      \centering
      \includegraphics[width=1.0\textwidth]{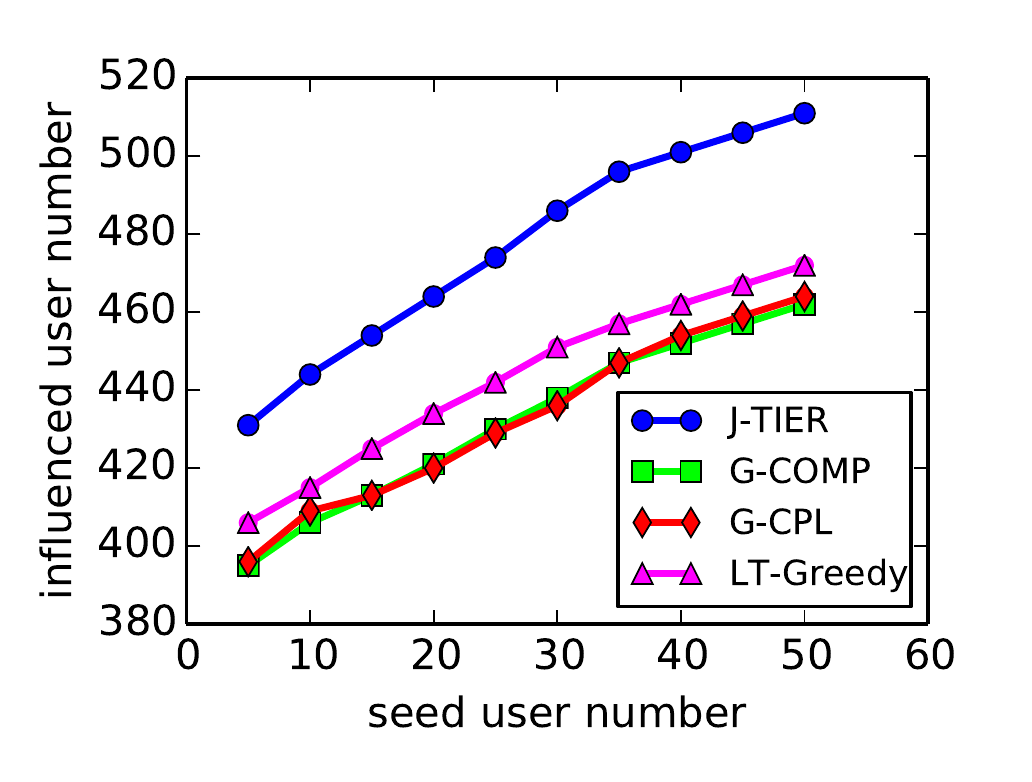}
    \end{minipage}
}
\subfigure[arXiv]{\label{eg_fig12_3}
    \begin{minipage}[l]{0.47\columnwidth}
      \centering
      \includegraphics[width=1.0\textwidth]{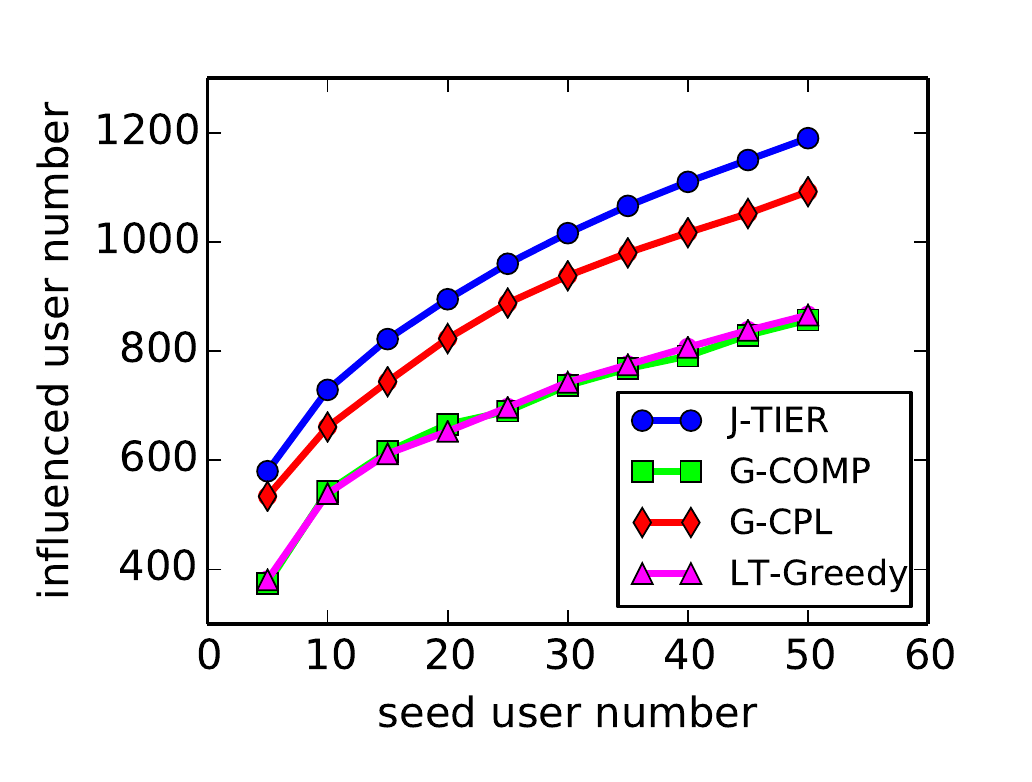}
    \end{minipage}
}
\subfigure[Epinions]{\label{eg_fig12_4}
    \begin{minipage}[l]{0.47\columnwidth}
      \centering
      \includegraphics[width=1.0\textwidth]{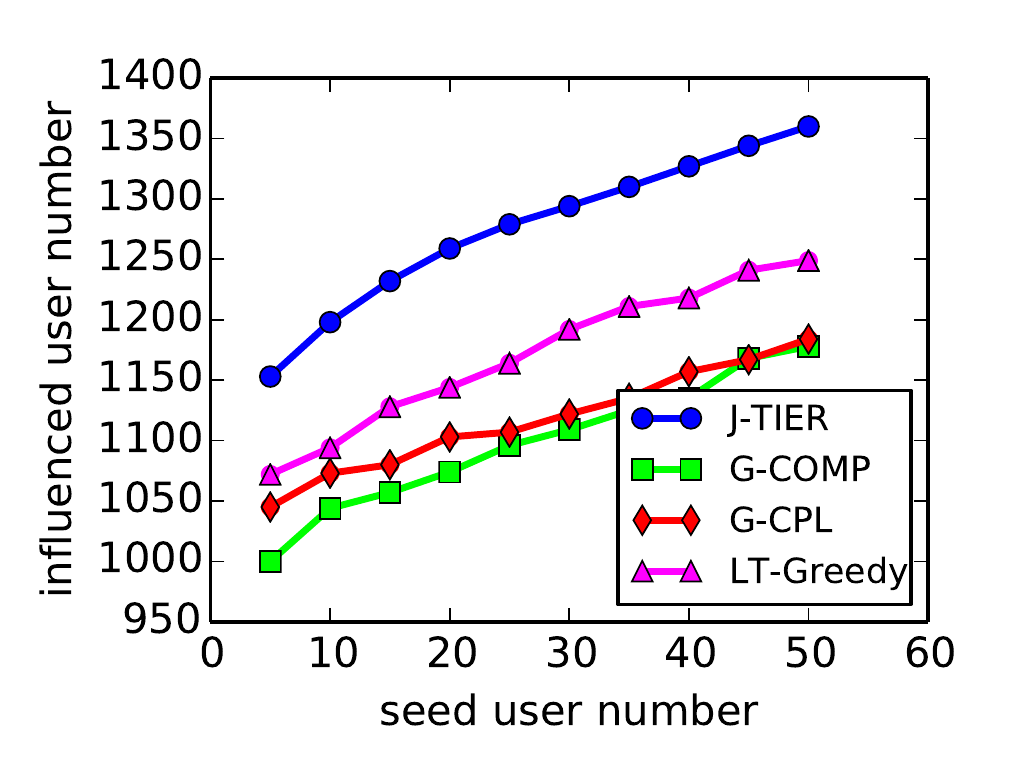}
    \end{minipage}
}
\caption{Experiment results of the {\problemunknown} problem.}\label{eg_fig12}
\end{figure*}

\begin{figure*}[t]
\centering
\subfigure[Facebook]{\label{eg_fig13_1}
    \begin{minipage}[l]{0.47\columnwidth}
      \centering
      \includegraphics[width=1.0\textwidth]{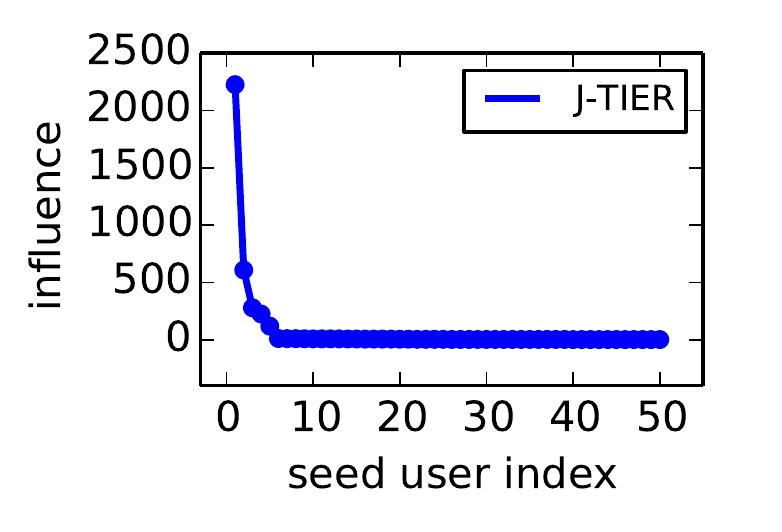}
    \end{minipage}
}
\subfigure[Wikipedia]{\label{eg_fig13_2}
    \begin{minipage}[l]{0.47\columnwidth}
      \centering
      \includegraphics[width=1.0\textwidth]{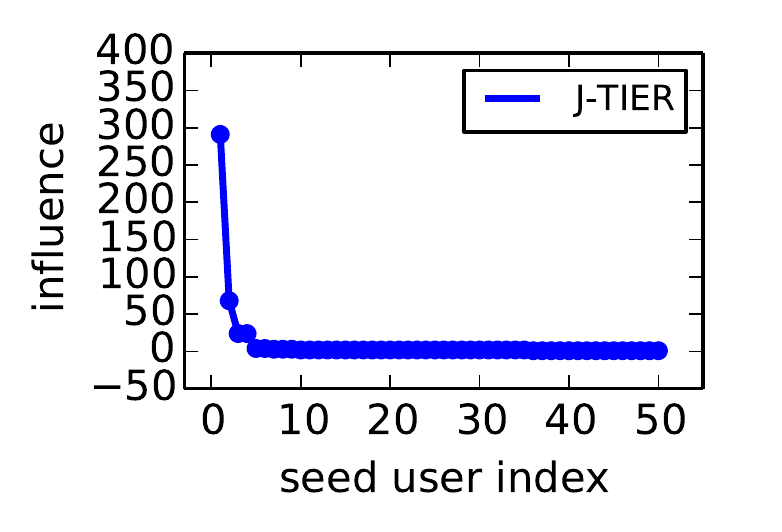}
    \end{minipage}
}
\subfigure[arXiv]{\label{eg_fig13_3}
    \begin{minipage}[l]{0.47\columnwidth}
      \centering
      \includegraphics[width=1.0\textwidth]{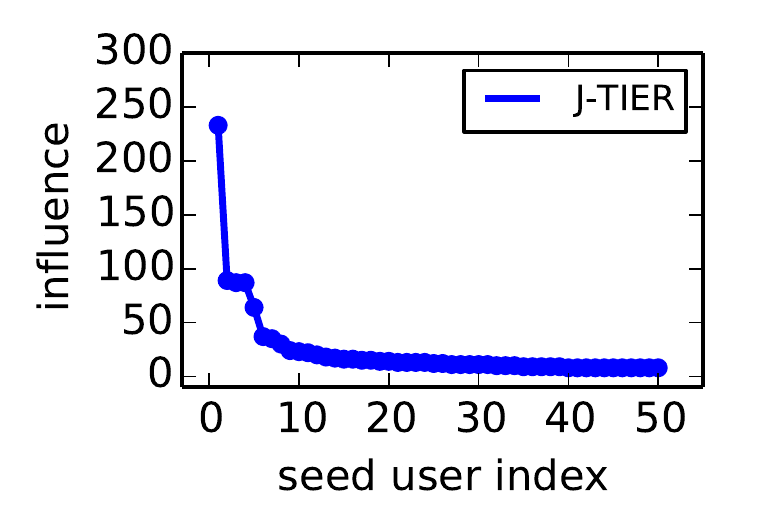}
    \end{minipage}
}
\subfigure[Epinions]{\label{eg_fig13_4}
    \begin{minipage}[l]{0.47\columnwidth}
      \centering
      \includegraphics[width=1.0\textwidth]{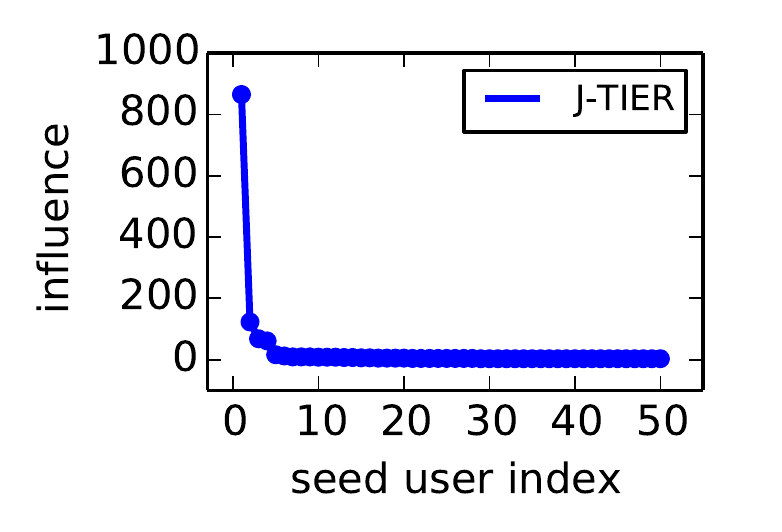}
    \end{minipage}
}
\caption{Influence achieved by each seed user selected by Algorithm {\ourunknown} in the {\problemunknown} problem.}\label{eg_fig13}
\end{figure*}

\subsection{Experiment Setting of the {\problemknown} Problem}

In this subsection, we will introduce comparison methods and experiment setups of the {\problemknown} problem.

\subsubsection{Comparison Methods}

In the {\problemknown} problem, the marketing strategies of all the other products are known in advance. ``\textit{Utilizing these known marketing strategies to select seed users for the target product can help achieve larger social influence in the social network}.'' To demonstrate such a claim, different methods are compared in the experiments, which can be divided into two categories:

\noindent \textbf{Methods using the known strategies}

\begin{itemize}

\item {\ourknown}: {\ourknown} based on the {\diffusion} diffusion model is the method proposed in this paper. Other products' known marketing strategies are used to update users thresholds towards the target product dynamically with the \textit{intertwined threshold updating strategy}. In each step, {\ourknown} selects the user who can lead to the maximum influence as the seed user.

\end{itemize}

\noindent \textbf{Methods without using the known strategies}

\begin{itemize}

\item {\LTGreedy}: {\LTGreedy} is the \textit{greedy} seed user selection method based on the traditional LT diffusion model. {\LTGreedy} ignores the existence of other products in seed user selection \cite{KKT03}.

\item {\LTPageRank}: {\LTPageRank} is based on the traditional LT diffusion model and doesn't use the know marketing strategies of other products. {\LTPageRank} is a heuristics-based method and chooses users with the top $K$ page rank scores as the final seed users \cite{BP98}.

\item {\LTInDegree}: {\LTInDegree} is quite similar to {\LTPageRank}: (1) it is a heuristics-based method, (2) it is based on traditional LT diffusion model, and (3) it doesn't use the known marketing strategies of other products. {\LTInDegree} chooses users with the top $K$ in degrees (i.e., \# followers) as the seed users \cite{CWY09}.

\item {\LTRandom}: {\LTRandom} chooses $K$ seed users from the network randomly from the network.

\end{itemize}

\subsubsection{Experiment Setup}

The connections among users in some networks are undirected, e.g., Facebook and arXiv, but in some others are directed, e.g., Wikipedia and Epinions. To unify different kinds of networks in our model, we replace undirected links, e.g., $u_i - u_j$, with two directed links $u_i \to u_j$, $u_i \gets u_j$, and links among users in our model are all directed. In the {\diffusion} diffusion model, each user can influence his neighbors with certain influence weights and has a threshold denoting the minimal required influence to be activated by other users. The weight of directed social link $(u_j \to u_i)$ ($u_j$ follows $u_i$ or $u_i$ influences $u_j$) quantifies the influence propagated from $u_i$ to $u_j$. In the experiments, the influence weight of link $(u_i, u_j)$ is quantified as $JC(u_i \to u_j) = \frac{\left | \Gamma(u_i) \cap \Gamma_{out}(u_j) \right |}{\left | \Gamma(u_i) \cup \Gamma_{out}(u_j) \right |}$, which is widely used in existing works \cite{ST11} and depends on not only the shared users between $u_i$ and $u_j$ but also the degrees of $u_i$ and $u_j$ respectively. Considering that there exist multiple products to be promoted in the network, for simplicity, the influence weights of link $(u_i \to u_j)$ in promoting different products are all set as $JC(u_i \to u_j)$. Meanwhile, users will have multiple thresholds towards all these products, which can be represented as $\mb{\theta}_i = (\theta_i^1, \cdots, \theta_i^n)$ and $\theta_i^j$ is the threshold of user $u_i$ towards product $p^j$. The thresholds are randomly selected from uniform distribution within range $[0, 1]$. In the experiment, we consider $4$ different products shown in Figure~\ref{fig:relationships}, where ``HP printer'' is the target product and ``Canon printer'', ``PC'' and ``Pepsi Diet'' are \textit{competing}, \textit{complementary} and \textit{independent} respectively to ``HP printer''. The \textit{threshold updating coefficient} between (1) \textit{independent} products is set as $1.0$; (2) \textit{competing} products is randomly selected from $[1, 2]$, and (3) \textit{complementary} products is randomly chosen from range $[0, 1]$.

The number of selected seed user for ``HP printer'' changes in range $\{5, 10, 15, \cdots, 45, 50\}$. For methods without utilizing the known strategies, we can just select seed users for ``HP printer'' based on the traditional LT model with methods {\LTGreedy}, {\LTPageRank}, {\LTInDegree} and {\LTRandom} without considering the other products, which is exactly how these methods work in traditional single-product problem settings.

Meanwhile, {\ourknown} will update the network with the \textit{intertwined threshold updating strategy} to use the known strategies of other products. The known seed users of products ``Canon printer'', ``PC'' and ``Pepsi Diet'' are selected with the {\LTGreedy} algorithm from the network, whose sizes are all $50$. The selected seed users of these products will propagate their influence in the network. Thresholds of users who get activated the products will be updated according to the \textit{threshold updating strategy}. Based on the updated network, we apply {\ourknown} to select seed users for ``HP printer''. 

To evaluate the performance of all these methods, we will calculate the number of users influenced by the seed users based on the updated network.

\subsection{Experiment Results of the {\problemknown} Problem}

The experiment results of different comparison methods are given in Figure~\ref{eg_fig11}, where Subfigures~\ref{eg_fig11_1}-~\ref{eg_fig11_4} correspond to Facebook, Wikipedia, arXiv and Epinions datasets respectively. 

Based on the results in Subfigures~\ref{eg_fig11_1}-~\ref{eg_fig11_4}, the number of users who get influenced generally increases as more seed users are selected for most methods except {\LTRandom}.  {\LTRandom} selects seed users randomly and the number of influenced users achieved by which can vary dramatically.



By comparing {\ourknown} with {\LTGreedy}, we observe that {\ourknown} can perform better than {\LTGreedy} consistently for different seed user set sizes in all these $4$ datasets. For example, in the arXiv dataset when seed user set size is $50$, the number of users get influenced by {\our} is $1,185$, which is over $50\%$ larger than the $753$ influenced users achieved by {\LTGreedy}. Experiments on other datasets show the similar results with various sizes of the seed users. It demonstrates that (1) the {\diffusion} diffusion model with \textit{threshold updating strategy} works better than the traditional LT model, and (2) utilizing the known \textit{marketing strategies} of other products can help lead to greater influence.


In Table~\ref{tab:intersection}, we show the intersections of seed user sets selected by different methods in different datasets, where the seed user set sizes are $50$. We can observe that (1) seed users selected by {\ourknown} have some overlaps with traditional greedy method {\LTGreedy}, (2) two heuristics-based methods {\LTPageRank} and {\LTInDegree} tend to select more common seed users, and (3) seed users selected by {\ourknown} is very different from those selected by {\LTPageRank}, {\LTInDegree} and {\LTRandom}. For example, in Facebook dataset, the intersection between seed user sets achieved by {\ourknown} and {\LTGreedy} is $31$ but those between {\ourknown} and other methods are $0$. So is the case in other datasets.

By comparing the performance of {\ourknown} with other comparison methods in all these $4$ networks, we observe that for densely connected networks (e.g., Facebook, Wikipedia, Epinions), heuristics (e.g., page rank, degree and greedy strategy) applied in traditional methods can work effectively. However, for sparse networks (e.g., arXiv), where the seed user selection problem will be more tough, these traditional heuristics no longer work well and the advantages of {\ourknown} are more obvious.

In sum, (1) \textit{threshold updating strategy} and the new {\diffusion} diffusion model works better than traditional LT model in addressing the {\ourknown} problem, (2) utilizing the known \textit{marketing strategies} of other products can help select better seed users, (3) seed users selected by {\ourknown} is quite different from those selected by other comparison methods, and (4) {\ourknown} can be applied to networks of different densities, especially the sparse/emerging networks \cite{ZY15}.


\subsection{Experiment Setting of the {\problemunknown} Problem}

In this subsection, we will introduce comparison methods and experiment setups of the {\problemunknown} problem.

\subsubsection{Comparison Methods}

In {\problemunknown} problem, the marketing strategies of other products are unknown and we consider the seed user selection process as a game among all the products. All products are assumed to be \textit{selfish} and want to choose users who can maximize their influence in the network. ``\textit{Meanwhile, in the seed user selection process, incorporating all the other products into the game can lead to better results}.'' To demonstrate such a claim, depending on the opponents incorporated in the game, the comparison methods used to address the {\problemunknown} problem can be divided into $2$ categories:

\noindent \textbf{Methods with Complete Game Opponents}

\begin{itemize}

\item {\ourunknown}: In seed user selection process, all the products (i.e., independent, competing and complementary products) are involved in the game. This is the {\ourunknown} method proposed in this paper.

\end{itemize}

\noindent \textbf{Methods with Partial Game Opponents}

\begin{itemize}

\item {\gamecompeting}: Enlightened by the analysis in \cite{TAM12}, we propose {\gamecompeting} (\underline{G}ame among \underline{COMP}eting products) as a potential comparison method, which can select seed nodes by only considering the competing products as the game opponents but ignoring the other two types of products.

\item {\gamecollaborative}: Method {\gamecollaborative} (\underline{G}ame among \underline{C}om\underline{PL}ementary products) extends the B-IMCP model proposed in \cite{NN12}, which can select seed nodes by only considering complementary products as the game opponents.

\item {\gameindependent}/{\LTGreedy}: Method {\gameindependent} (\underline{G}ame among \underline{INDEP}endent products) ignores the \textit{competing} and \textit{complementary} products and only considers the \textit{independent products} as the potential game opponents. Considering that \textit{independent products} will not change users' thresholds towards the target product, method {\gameindependent} is identical to the traditional \textit{step-wise greedy method} {\LTGreedy}, which ignores all the other products in the network \cite{KKT03}.

\end{itemize}

\subsubsection{Experiment Setup}

The experiment setup of the {\problemunknown} problem is similar to that of the {\problemknown} problem. For different comparison methods, specific types of products are involved in the game and the selected seed users at each step are recorded. In evaluation, we simulate the game among different products again, where seed users of other products are those selected by {\ourunknown} but seed users of the target product are replaced with those selected by different comparison methods. In the simulation, each product choose its seed users by turns and the influence of the seed users will propagate within the network and update users' thresholds right after it is selected. We calculate the number of users get influenced by the seed users of the target product to evaluate the comparison methods' performance.

\subsection{Experiment Results of the {\problemunknown} Problem}

The results of different comparison methods in addressing the {\problemunknown} problem on different datasets are available in Figure~\ref{eg_fig12}, where Subfigures~\ref{eg_fig12_1}-~\ref{eg_fig12_4} correspond to the Facebook, Wikipedia, arXiv and Epinions networks respectively. 

Based on Subfigures~\ref{eg_fig12_1}-~\ref{eg_fig12_4}, the results achieved by {\ourunknown} is much better than those obtained by other methods. It shows that for the target product, when selecting seed users, considering all the existing products as game opponents (including competing, complementary and independent products) can help make better choices. For example, in Epinions network when seed user set size is $50$, the influenced user numbers achieved by {\ourunknown}, {\gamecompeting}, {\gamecollaborative} and {\LTGreedy} are $1390$, $1178$, $1184$ and $1249$ respectively. The results achieved by considering all the products in the game is (1) $11.2\%$ better than that achieved by only considering \textit{competing products} in the game, (2) $17.3\%$ better than that gained by considering \textit{complementary products} only, and (3) $7.28\%$ better than that obtained by considering independent products only.

In addition, in Figure~\ref{eg_fig13}, we show the influence introduced by each seed user selected by {\ourunknown} in addressing the {\problemunknown} problem in the $4$ different networks. We observe that the new influence introduced by the new seed users will decrease dramatically as the number of existing seed users increases. The first seed user generally can lead to the largest amount of influence to the network.

In sum, making full considerations of all these three types of products opponents in the game can help make more comprehensive and better seed user selections in viral marketing.


\section{Related Work} \label{sec:relatedwork}

Viral marketing (i.e., influence maximization) problem in customer networks first proposed by Domingos et al. \cite{DR01} has been a hot research topic. Richardson et al. \cite{RD02} study the viral marketing based on knowledge-sharing sites and propose a new model which needs less the computational cost than the model proposed in \cite{DR01}. Kempe et al. propose to study the influence maximization problem through a social network \cite{KKT03} and propose to different diffusion models: Independent Cascade (IC) model and Linear Threshold (LT) model, which have been widely used in later influence maximization papers.

The approximation methods proposed in \cite{KKT03} to address the influence maximization problem can be very time consuming. To address such a problem, lots of works have been done to reduce computation cost. Leskovec et al. \cite{LKGFVG07} propose an efficient methods which can achieve near optimal results but be $700$ times faster than a simple greedy method. Chen et al. \cite{CWY09} propose to study the efficient influence maximization problem from two complementary directions: (1) improve the greedy method; (2) propose new degree discount heuristics to improve the spread. Chen et al. \cite{CWW10} propose a heuristic algorithm which is easily scalable to millions of nodes and edges.

In recent years, various variants of the influence maximization problem have been proposed. Song et al. \cite{SCHT07} propose to identify the opinion leaders in bolgosphere. Goyal et al. \cite{GBL08} study the leader discovery problem from the community actions. Hartline et al. \cite{HMS08} study the optimal marketing strategy design problem over social networks to maximize the revenue instead. Agarwal et al. \cite{ALTY08}  propose to identify the influential bloggers in a community. Provost et al. \cite{PDHZM09} propose to select audience for online brand advertising. Zhan et al. propose to extend the traditional single-network viral marketing problem to multiple aligned networks in \cite{Zhan2015}.


Meanwhile, the promotions of multiple products can exist in social networks simultaneously, which can be independent, competing or complementary. Datta et al. \cite{DMS10} study the viral marketing for multiple independent products at the same time and aim at selecting seed users for each products to maximize the overall influence. Pathak et al. \cite{PBS10} propose a generalized linear threshold model for multiple cascades. Bharathi et al. \cite{BKS07} propose to study the competitive influence maximization in social networks, where multiple competing products are to be promoted. He et al. \cite{HSCJ12} propose to study the influence blocking maximization problem in social networks with the competitive linear threshold model. Carnes et al. \cite{CNWZ07} study the influence maximization problem in a competitive social network from a follower's perspective and Chen et al. \cite{CCCKLRSWWY11} study the influence maximization in social networks when negative opinions can emerge and propagate. Multiple threshold models for competitive influence in social networks are proposed in \cite{BFO10}, whose submodularity and monotonicity are studied in details. A nash equilibrium based model is proposed by Dubey et al. \cite{DGM06} to compete for customer in online social networks. Meanwhile, Narayanam et al. \cite{NN12} study the viral marketing for product cross-sell through social networks to maximize the revenue, where products can have promotion cost, benefits and promotion budgets.

\section{Conclusion}\label{sec:conclusion}

\begin{figure*}[t]
\centering
\subfigure[counter examples of \textit{monotone} property]{ \label{fig_eg4_1}
    \begin{minipage}[l]{0.95\columnwidth}
      \centering
      \includegraphics[width=0.8\textwidth]{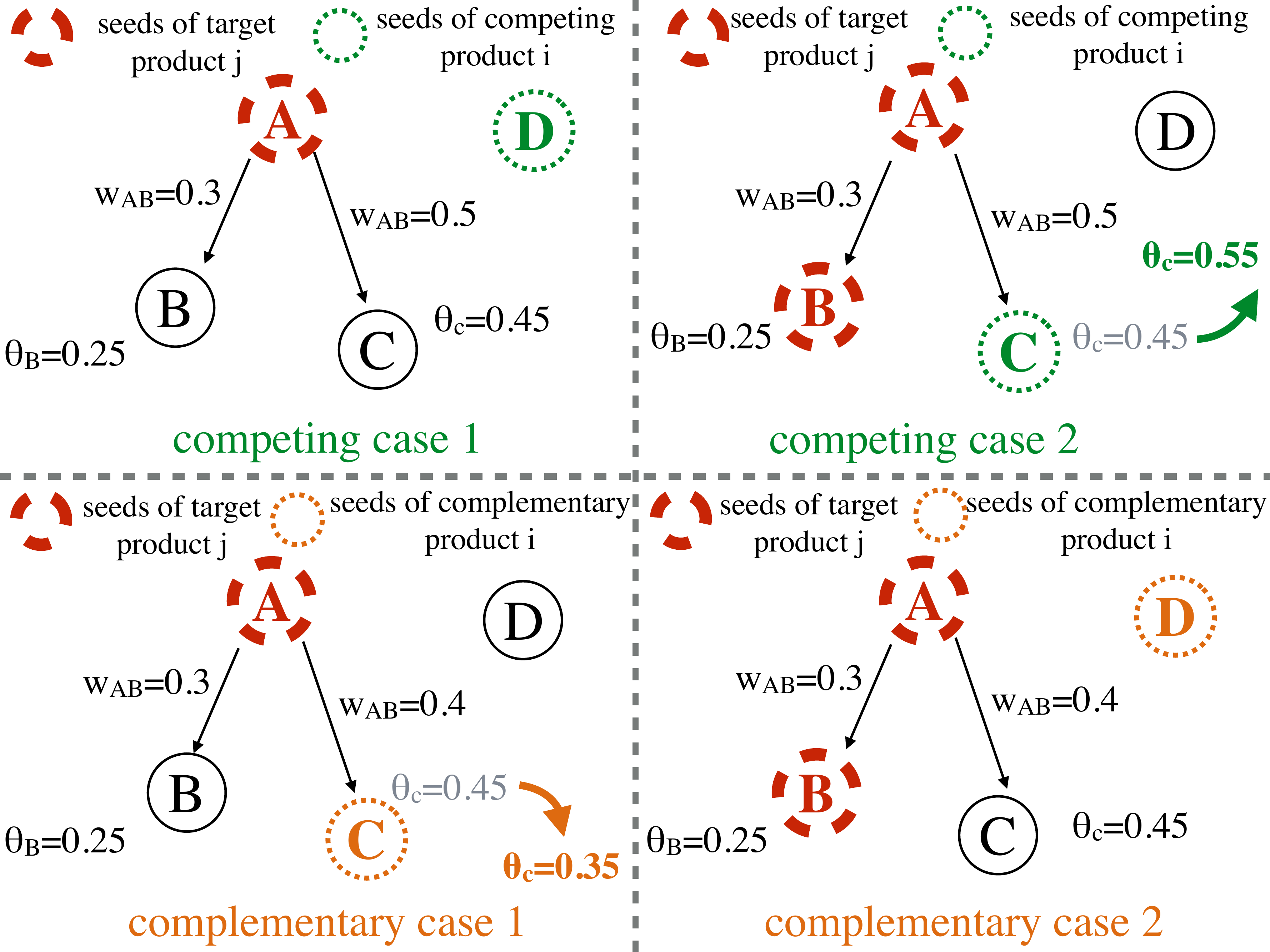}
    \end{minipage}
}
\subfigure[counter examples of \textit{submodular} property]{ \label{fig_eg4_2}
    \begin{minipage}[l]{0.95\columnwidth}
      \centering
      \includegraphics[width=0.8\textwidth]{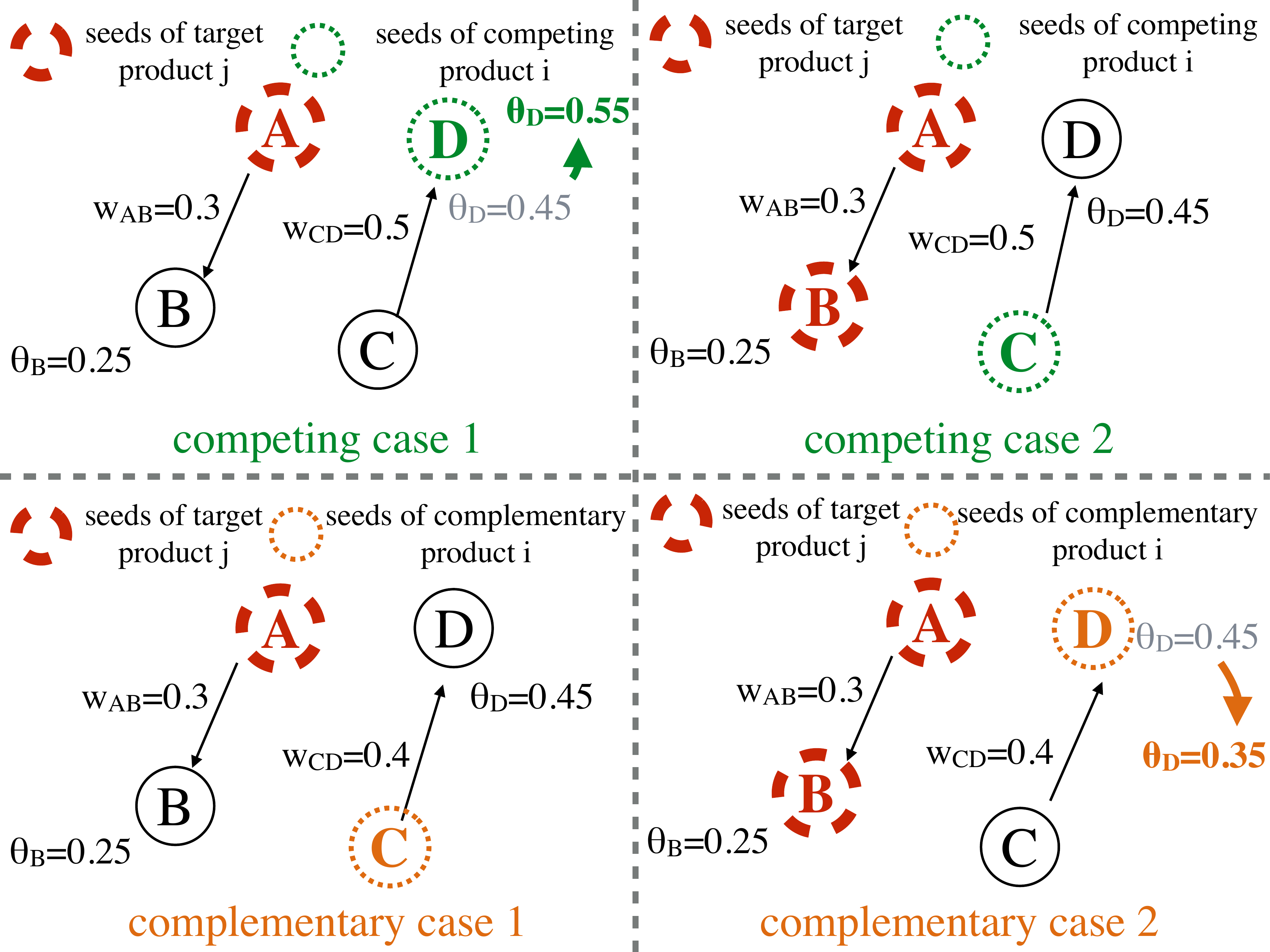}
    \end{minipage}
}
\caption{Counter examples of monotone and submodular properties.}\label{fig_eg4}
\end{figure*}

In this paper, we have studied  the {\problem} problem in online social networks. A novel unified framework {\our} has been proposed to address the {\problem} problem. {\our} is based on a novel diffusion model {\diffusion}, which can update users' thresholds dynamically. For the {\problemknown} problem, \textit{greedy} method {\ourknown} selects the optimal seed users at each step and can achieve a $(1-\frac{1}{e})$-approximation to the optimal results. For the {\problemunknown} problem, {\ourunknown} formulates the seed user selection process of multiple products as a game and selects the optimal seed users step by step based on the inferred \textit{marketing strategies} of other products. Extensive experiments on 4 real-world social network datasets demonstrate the superior performance of {\ourknown} and {\ourunknown} in addressing the {\problemknown} and {\problemunknown} problems.

\balance
\bibliographystyle{plain}
\bibliography{reference}
\label{sec:appendix}
\section{Appendix}
\small

\vspace{-1ex}
\subsection{Appendix-A}

\noindent Theorem 5: Based on the {\diffusion} diffusion model, the \textit{joint influence function} is \textit{monotone} and \textit{submodular} if all the other products are \textit{independent} to $p^j$.

\noindent \textit{Proof}: If all the other products are \textit{independent} to product $p^j$, according to the \textit{threshold updating strategy}, users' thresholds to the target product $p^j$ will not be updated by these products. As a result, the {\problem} problem identical to the \textit{traditional single-product viral marketing} problem, which has been proved to be \textit{submodular} and \textit{monotone} in \cite{KKT03}.


\vspace{-1ex}
\subsection{Appendix-B}

\noindent Theorem 6: Based on the {\diffusion} diffusion model, the \textit{joint influence function} is not \textit{monotone} if there exist products which are either \textit{competing} or \textit{complementary} to the target product $p^j$.

\noindent \textit{Proof}: Similar to \cite{BFO10}, we propose to prove Theorem~\ref{th:not_monotone} with potential counter examples shown in Figure~\ref{fig_eg4_1}, where we can find one product $p^i$ to be either \textit{competing} or \textit{complementary} to $p^j$.

\noindent \textit{Case (1)}: \textit{competing products exist}: as shown in the upper two plots in Figure~\ref{fig_eg4_1}, we have $4$ users in the network $\{A, B, C, D\}$ and we want to select seed users for products $p^i$ and $p^j$. The influence from $A$ to $B$ and $C$ are $0.3$ and $0.5$, whose original thresholds to the target product $p^j$ are $0.25$ and $0.45$ respectively. In the example, the \textit{seed users} selected for two \textit{competing} products $p^j$ and $p^i$ are (1) $\{A\}$ and $\{D\}$ respectively in competing case 1 at the upper left corner; and (2) $\{A, B\}$ and $\{C\}$ in competing case 2 at the upper right corner. In competing case 1, $p^j$ can influence $3$ users $\{A, B, C\}$ as the influence from $A$ to $B$ and $C$ can both exceed their thresholds, i.e., $I(\mathcal{S}^j = \{A\}; \mathcal{S}^i = \{D\}) = 3$. However, in competing case 2, $p^j$ can only influence $2$ users, even though the seed user set has been expanded by adding $B$ as a \textit{seed user}, i.e., $I(\mathcal{S}^j = \{A, B\}; \mathcal{S}^i = \{C\}) = 2$. The reason is that the competing product $p^i$ selects $C$ as the seed user which increase $C$'s threshold towards $p^j$ from $0.45$ to $0.55$. So, we can find a counter example where $\{A\} \subset \{A, B\}$ but $I(\mathcal{S}^j = \{A\}; \mathcal{S}^i = \{D\}) > I(\mathcal{S}^j = \{A, B\}; \mathcal{S}^i = \{C\})$, when there exists \textit{competing} product $p^i$ in the network.

\noindent \textit{Case (2)}: \textit{complementary products exist}: similar counter example are shown in the lower two plots of Figure~\ref{fig_eg4_1}, which are identical to the upper two plots except that the influence from $A$ to $C$ for product $p^j$ is changed to $0.4$ and $p^i$ is \textit{complementary} to $p^j$ instead. In complementary case 1, $p^i$ selects $C$ as the seed user, which can decrease $C$'s threshold towards $p^j$ and $p^j$ can achieve a influence of $3$ by choosing $A$ as the seed user. However, in complementary case 2, $p^i$ selects $D$ as the seed user and $p^j$ can only influence $2$ users even though the seed user set has been expanded by adding $B$ to the set. So, we can find a counter example where $\{A\} \subset \{A, B\}$ but $I(\mathcal{S}^j = \{A\}; \mathcal{S}^i = \{C\}) > I(\mathcal{S}^j = \{A, B\}; \mathcal{S}^i = \{D\})$ when there exists \textit{complementary} product $p^i$ in the network.
\vspace{-1ex}
\subsection{Appendix-C}

\noindent Theorem 7: For the {\diffusion} diffusion model, the \textit{joint influence function} is not \textit{submodular} if these exist products which are either \textit{competing} or \textit{complementary} to $p^j$.


%
\noindent \textit{Proof}: We propose to prove Theorem~\ref{th:not_submodular} with potential counter examples shown in Figure~\ref{fig_eg4_2}, where we can find one product $p^i$ to be either \textit{competing} or \textit{complementary} to $p^j$. 

\noindent \textit{Case (1)}: \textit{when competing products exist}: Let $\mathcal{T} = \{A\} \subset \mathcal{S} = \{A, B\}$ and $u = C$. In the competing case 1, $\mathcal{T}$ is the seed user set selected by product $p^j$ and $\{D\}$ is selected as the seed user by product $p^i$, which increase $D$'s threshold to $p^j$ from $0.45$ to $0.55$. As a result, $p^j$ can only influence 2 users ($\{A, B\}$) when using $\mathcal{T}$ as the seed user set and influence $3$ users ($\{A, B, C\}$) when using $\mathcal{T} \cup \{u\}$ as the seed user set. However, in the competing case 2, where $p^i$ selects $C$ as the seed user, $p^j$ can activate $2$ users ($\{A, B\}$) when using $\mathcal{S}$ as the seed user set but can activate $4$ users ($\{A, B, C, D\}$) when using $\mathcal{S} \cup \{u\}$ as the seed user set. So, we can find a counter example where $\mathcal{T} = \{A\} \subset \mathcal{S} = \{A, B\}$ and $u = C$, but $I(\mathcal{S}^j = \mathcal{T} \cup \{u\}; \mathcal{S}^i = \{D\}) - I(\mathcal{S}^j = \mathcal{T}; \mathcal{S}^i = \{D\}) < I(\mathcal{S}^j = \mathcal{S} \cup \{u\}; \mathcal{S}^i = \{C\}) - I(\mathcal{S}^j = \mathcal{S}; \mathcal{S}^i = \{C\})$.

\noindent \textit{Case (2)}: \textit{when complementary products exist}: similar counter example is shown in the lower two plots of Figure~\ref{fig_eg4_2}, where $p^i$ is \textit{complementary} to $p^i$. We can also find a counter example where $\mathcal{T} = \{A\} \subset \mathcal{S} = \{A, B\}$ and $u = C$, and $I(\mathcal{S}^j = \mathcal{T} \cup \{u\}, \mathcal{S}^i = \{C\}) - I(\mathcal{S}^j = \mathcal{T}, \mathcal{S}^i = \{D\}) < I(\mathcal{S}^j = \mathcal{S} \cup \{u\}, \mathcal{S}^i = \{C\}) - I(\mathcal{S}^j = \mathcal{S}, \mathcal{S}^i = \{D\})$. As a result, For the {\diffusion} diffusion model, the \textit{joint influence function} is not \textit{submodular} if these exist products which are either \textit{competing} or \textit{complementary} to $p^j$.

\end{document}